\documentclass[printer]{aa}
\usepackage{natbib}
\bibpunct{(}{)}{;}{a}{}{,} 
\usepackage{graphicx}
\usepackage{txfonts}
\usepackage[bookmarks]{hyperref}

\begin{document}


\title{Lepto-hadronic model for the broadband emission of Cygnus X-1}
\titlerunning{Lepto-hadronic model for the broadband emission of Cygnus X-1}

\author{Carolina Pepe \inst{\ref{inst1}} \and  Gabriela S. Vila \inst{\ref{inst1}} \and Gustavo E. Romero \inst{\ref{inst1}, \ref{inst2}}}
\institute{Instituto Argentino de Radioastronom\'ia (IAR-CONICET), C. C. 5, (1894) Villa Elisa, Buenos Aires, Argentina \\
\email{carolina@iar.unlp.edu.ar}\label{inst1}
\and Facultad de Ciencias Astronómicas y Geofísicas, Universidad Nacional de La Plata (FCAG-UNLP)\label{inst2}, Paseo del Bosque S/N, (1900) La Plata, Buenos Aires, Argentina}


\abstract{Cygnus X-1 is a well observed microquasar. Broadband observations at all wavelengths have been collected over the years. The origin of the MeV tail observed with \emph{COMPTEL} and \emph{INTEGRAL} is still under debate and it has mostly been attributed to the corona, although its high degree of polarization suggests it is synchrotron radiation from a jet. The origin of the transient emission above $\sim 100$ GeV is also unclear.}
{We aim to disentangle the origin of the broadband spectral energy distribution (SED) of Cygnus X-1, focusing particularly on the gamma-ray emission, and to gain information on the physical conditions inside the jets. }
{We develop and apply a lepto-hadronic, inhomogeneous jet model to the non-thermal SED of Cygnus X-1. We calculate the contributions to the SED of both protons and electrons accelerated in an extended region of the jet. We also estimate the radiation of charged secondaries produced in hadronic interactions, through several radiative processes. Absorption effects are considered. We produce synthetic maps of the jets at radio wavelengths.}
{We find two sets of model parameters that lead to good fits of the SED. One of the models fits 
all the observations, including the MeV tail. This model also predicts hadronic gamma-ray emission slightly below the current upper limits. The flux predicted at \mbox{8.4 GHz} is in agreement with the observations available in the literature, although the synthetic source is more compact than the imaged radio jet.}
{Our results show that the MeV emission in Cygnus X-1 may be jet synchrotron radiation. This depends mainly on the strength of the jet magnetic field and the location of the injection region of the relativistic particles. Our calculations show that there must be energetic electrons in the jets quite far from the black hole.}

\maketitle

\section{Introduction}

Microquasars (MQs) are X-ray binaries that exhibit collimated, mildly relativistic outflows. Among all galactic X-ray binaries, Cygnus X-1 is the strongest candidate to host a black hole. This source has been the target of extensive monitoring campaigns that allowed to estimate the parameters of the binary and to obtain detailed spectra at all wavelengths. Cygnus X-1 is located at $1.86~\rm{kpc}$ from Earth \citep{reid2011}. A high-mass stellar companion of spectral type O9.7 Iab and mass \mbox{$\sim 20~M_{\odot}$} and a black hole of \mbox{$14.8~M_{\odot}$} \citep{orosz2011} form the binary system. 
 
In the X-ray band Cygnus X-1 switches between the typical low/hard and high/soft states of X-ray binaries. The high/soft state is characterized by a blackbody component of \mbox{$kT \lesssim 0.5~\rm{keV}$} from an accretion disk, and a soft power-law with spectral index \mbox{$\Gamma \sim 2 -3$}. The source, however, spends most of the time in the low/hard state, in which the spectral energy distribution is well described by a power-law of spectral index \mbox{$\Gamma \sim 1.7$} that extends up to a high-energy cutoff at \mbox{$\sim 150$ keV} \citep[e.g.][]{dove-etal1997, poutanen1998}. The origin of this power-law is the  Comptonization of disk photons by thermal electrons in a hot corona that partially covers the inner region of the disk. The detection of a Compton reflection bump and the Fe K$\alpha$ line  at $\sim6.4$ keV support the presence of the corona during the low/hard state.   Additionally, intermediate spectral states have also been reported \citep{belloni1996}. 

Persistent and transient jets have been resolved at radio wavelengths in Cygnus X-1 during the low/hard state \citep{stirling2001, fender2006, pandey2006, Rushton-etal2011}.\footnote{There is also some evidence of a jet-like, compact, unresolved structure during the high/soft state, see \cite{Rushton-etal2012}.} The outflow is extremely collimated \citep[aperture angle $\sim 2^\circ$,][]{stirling2001} and propagates at an angle of $\sim 29^\circ$ with the line of sight \citep{orosz2011}. The radio emission is modulated by the orbital period of the binary because of absorption in the wind of the companion star \citep{brocksopp2002, lachowicz-etal2006, zdziarski2012b}.

Cygnus X-1 is one of the two confirmed MQs that is a gamma-ray source.\footnote{The other one is Cygnus X-3. }  The first detection of soft gamma rays up to a few MeV was achieved with the instrument \emph{COMPTEL} aboard the \emph{Compton Gamma-Ray Observatory} \citep{mcConnell2000, mcConnell2002}. Emission in the same energy range was later observed with \emph{INTEGRAL} \citep{cadolleBell-etal2006, laurent2011, jourdain2012}. The  \emph{INTEGRAL} detections represented a breakthrough since it was found that the \mbox{$\sim 400$ keV - 2 MeV} photons were highly polarized  \citep{laurent2011, jourdain2012, rodriguez2015}. 

At higher energies Cygnus X-1 is fundamentally a transient source on timescales of 1-2 d. Episodes of gamma-ray emission have been detected with the satellite \emph{AGILE} between 100 MeV and a few GeV in the low/hard state (\citealt{sabatini2010}, see also \citealt{bulgarelli-etal2010}) and marginally during the low/hard-to-high/soft transition \citep{sabatini2013}. The analysis of \emph{Fermi}-Large Area Telescope (LAT) data at 0.1-10 GeV revealed weak flares (three of them quasi-simultaneous with \emph{AGILE} detections; \citealt{bodaghee-etal2013}, see also \citealt{TamYang2015}) and weak steady emission \citep{malyshev2013}.

Finally, Cygnus X-1 has been observed in the very high energy band ($\geq 100$ GeV) with the Major Atmospheric Gamma Imaging Cherenkov (MAGIC) telescope  during the low/hard state \citep{albert2007}. During inferior conjunction a flare (duration of less than a day, rising time $\sim 1$ h) was likely detected with a significance of  $4.0\sigma$  ($3.2\sigma$ after trial correction).  Only upper limits could be obtained for the steady emission. 

The broadband spectral energy distribution (SED) of Cygnus X-1 in the low/hard state displays several components. The radio emission is synchrotron radiation of relativistic electrons accelerated in the jets. This component is observed up to its turnover at infrared frequencies, where the stellar continuum takes over. The emission of the disk/corona dominates up to the hard X-rays, but the origin of the MeV tail observed with \emph{COMPTEL} and \emph{INTEGRAL} is still disputed. Its high degree of polarization suggests this component is emitted in an ordered magnetic field such as is expected to exist in the jets, a result supported by the fact that the polarized X-ray emission is only clearly detected during the low/hard state \citep{rodriguez2015}. In this scenario the MeV tail would be the cutoff of the jet synchrotron spectrum, see for example the fits to the data obtained by \cite{rahoui2011}, \cite{malyshev2013}, \cite{RussellShahbaz2014}, \cite{zdziarski2012}, \cite{zdziarski2014}, and \cite{zhang2014}
.  An alternative model was introduced by \cite{romero2014}, where the MeV tail is synchrotron radiation of secondary non-thermal electrons in the corona. This model predicts significant polarization also during intermediate spectral states, something that cannot be presently ruled out from the data \citep{rodriguez2015}.

All known gamma-ray binaries host a high-mass donor star, a fact that points to a fundamental role played by the stellar wind and/or  radiation field in the mechanisms that produce the high-energy photons. In leptonic models for MQs gamma rays are produced by inverse Compton (IC) scattering of the stellar radiation off relativistic electrons \citep{boschRamon2006, khangulyan2008}, whereas in hadronic models gamma rays are generated by the decay of neutral pions injected in the interactions of non-thermal protons in the jets with cold protons of the stellar wind \citep{romero2003, romero-etal2005}.  

The inclusion of relativistic protons in the jets brings about a feature absent in purely leptonic models, namely the production of secondary particles (neutrinos, electron-positron pairs, muons, pions) in high-energy hadronic interactions. The cooling of charged secondaries may contribute to the radiative spectrum of the jets. Neutrinos are a by-product unique to hadronic interactions, and their detection would definitely settle the question of the composition of relativistic jets; see for example \cite{LevinsonWaxman2001}, \cite{bednarek2005} and \cite{reynoso2009}. Up to date, hadrons (specifically iron nuclei) have only been detected in two sources: SS 433 \citep{migliari2002} and 4U 1630C47 \citep{diazTrigo2013}.  Nevertheless, given the correlation between accretion and ejection observed in MQs \citep{mirabel1998} it is reasonable to assume that the outflows have a composition similar to that of the accretion flow. Furthermore, in Cygnus X-1 the effects of the impact of the jets in the interstellar 
medium suggest that they carry a significant amount of kinetic energy in cold protons \citep{gallo2005, Heinz2006, sell2015}.

In this work we apply a lepto-hadronic, inhomogeneous jet model to study the broadband SED of Cygnus X-1. The model is based on previous works by \cite{romero-vila2008} and \cite{vila-etal2012} and their applications to low-mass microquasars \citep{vila2010}. Here we present an extended version of the model that also accounts for the interaction between the relativistic particles in the jets and the wind and radiation of the donor star. We consider a lepto-hadronic jet, i.e. one with a content of both non-thermal electrons and protons. Although we do not deal with the mechanism that accelerate these particles, we assume that it operates over an extended region of the jet with varying physical parameters (magnetic field, density of the internal and external radiation and matter  fields, etc.). We account for the cooling of the non-thermal particles and their transport along the jet by convection. With the complete characterization of the particle distributions in energy and space, we compute all the non-thermal components of the broadband SED of the jet - including the contributions of both primary and secondary charged particles - and assess the effects of absorption in the stellar radiation field. Two best-fit SEDs reproduce the available multi-wavelength observations of Cygnus X-1 from radio to gamma rays. For each of them we calculate synthetic radio maps of the jet to be compared with actual interferometric images of the source. From the combined analysis of the SED and the radio maps, we are able to draw some conclusions on the hadronic content of the jet, the behaviour of the magnetic field along the jet, and the possible sites of particle acceleration. In this regard our model goes beyond and complements other available models for the non-thermal SED of \mbox{Cygnus X-1} such as those of \cite{zdziarski2012} and \cite{zdziarski2014}, providing new tools that may help to gain insight into the physical conditions in the jets of microquasars.

The article is organized as follows. In Section \ref{sec:model} we present the basics of the model and discuss the processes that contribute to the
acceleration and cooling of relativistic particles. In Section \ref{sec:results} we present the results of the application of the model to Cygnus X-1: cooling rates, 
particle distributions, best-fit SEDs and radio maps. In Section \ref{sec:discussion} we discuss the results and the perspectives they open for future investigations. 

\section{The model}
\label{sec:model}

\subsection{Basic scenario}

\begin{figure}
 \includegraphics[width = 0.4 \textwidth]{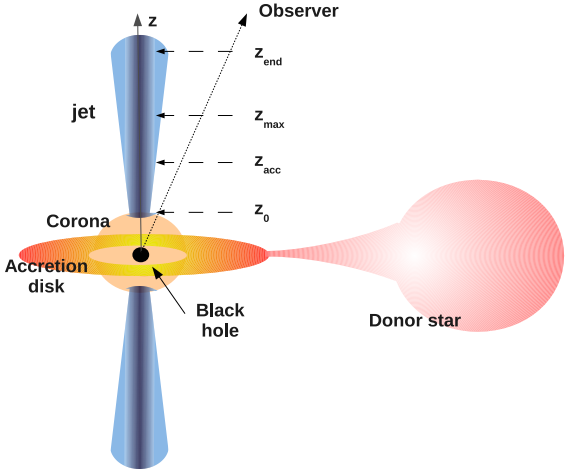}
 \caption{Basic sketch of the binary and the jet (not to scale). }
 \label{fig:sketch}
\end{figure}

In this section we summarize the main features of the jet model. For an extensive description, details, and all relevant formulae the reader is referred
to \cite{vila-etal2012} and references therein. A basic sketch of the binary system and the jet is shown in Fig. \ref{fig:sketch}. The massive star is
located at a distance $a_\bigstar = 3.2 \times 10^{12}~\rm{cm}$ from the black hole and injects matter and photons in the medium through its wind and its radiation field, respectively.  In all  our calculations
the binary is assumed to be at the superior conjunction; the impact of this hypothesis is discussed in Section \ref{sec:discussion}.  The black hole accretes matter from the strong stellar wind.\footnote{However, Cygnus X-1, unlike others HMMQs, is very close to filling its Roche lobe, and there are occasional episodes of accretion through Roche lobe overflow.} An accretion disk
extends from an inner radius $R_{\rm in}\sim 5 \times 10^7 ~\rm{cm}$ to an outer radius $R_{\rm out}\sim 2 \times 10^{11}~\rm{cm}$; it is modeled as a standard thin disk \citep{ShakuraSunyaev1973}. Closer to the black hole, the plasma inflates 
to form a hot corona. The hard X-ray emission from the corona is as a power-law of spectral index $\alpha$ with an exponential cutoff at photon energy $\epsilon = \epsilon_c$,

\begin{equation}
 n_{\rm ph} \propto  \epsilon^{-\alpha} \exp{\left(-\frac{\epsilon}{\epsilon_c}\right)};
\label{eq:corona_spectrum}
\end{equation}

\noindent we adopt $\alpha = 1.6$ and $\epsilon_c = 150$~keV \citep{poutanen1997}. A pair of symmetrical, conical jets (with a half-opening angle $\theta_{\rm op}$) is launched perpendicularly to the 
accretion disk at a distance $z_0$ from the black hole and propagates with constant bulk Lorentz factor $\Gamma_{\rm{jet}}$ up 
to $z_{\rm{end}}$.\footnote{We define the $z$-axis along the jet axis. Radial symmetry is assumed: the model parameters depend only on the coordinate $z$.} 
The jet axis forms an angle $\theta_{\rm{jet}}$ with the direction of the line of sight. Each jet carries a power $L_{\rm{jet}}$. Equipartition between the magnetic and the bulk kinetic energy densities is assumed at the jet base. This  allows to estimate the value of $B_0 = B(z_0)$.

Electrons and protons in the jets are accelerated via a diffusive mechanism mediated by shocks. A total power

\begin{equation}
 L_{\rm{rel}} = q_{\rm{rel}} L_{\rm{jet}} \qquad q_{\rm{rel}} < 1
\end{equation}

\noindent is transferred to the relativistic particles; $L_{\rm{rel}}$ is in turn the sum of the powers injected in relativistic electrons and protons, that in our model are related as

\begin{equation}
 L_{\rm{p}} = a L_{\rm{e}}.
\end{equation}

\noindent The value of the parameter $a$ determines the energetically dominant non-thermal component of the jet (equipartition for $a = 1$, proton-dominated for $a > 1$, and electron-dominated otherwise).

\subsection{Radiative processes and particle cooling}
  \label{sec:rates}
Relativistic particles lose energy through several processes. For any given mechanism the cooling rate is defined as 

\begin{equation}
 t^{-1} = -\frac{1}{E} \frac{dE}{dt},
\end{equation}

\noindent where $E$ is the particle energy. Since there must be an external medium to confine the outflows, particles lose energy adiabatically by exerting work on the walls of the jet, as well as radiatively. \footnote{All cooling rates are calculated in the jet co-moving reference frame except for the case of proton-proton collisions, that is calculated in the observer frame and then transformed.} 

Leptons (both primary and secondary) cool radiatively via synchrotron, relativistic Bremsstrahlung and inverse Compton emission. We assume three different target fields for IC scattering: synchrotron radiation from primary electrons 
(synchrotron self-Compton, SSC), X-ray photons radiated from the accretion disk (IC-disk), and stellar blackbody photons (blackbody Compton, BBC).  We calculated the BCC cooling rate in the full Klein-Nishina regime as in \cite{khangulyan2014}; for the rest of the processes the formulas are given in \cite{vila-etal2012} and references therein.  The star is assumed to radiate as a blackbody of $T_{\bigstar} = 2.8 \times 10^4$ K. Since the stellar photon distribution is anisotropic as seen from the jet
frame, the full angle-dependent IC cross section must be applied. The expression for this cross section in the head-on approximation is given for example in \cite{dermer1993}.

Protons cool via synchrotron radiation, proton-proton ($pp$) and proton-photon interactions ($p\gamma$). The targets for $pp$ collisions are the thermal protons in the jet and in the stellar wind ($pp$-star), whereas the photons for $p\gamma$ interactions are provided by the synchrotron field of primary electrons and the stellar radiation.   Hadronic interactions inject pions in the jet. Neutral pions subsequently decay producing two photons, 

\begin{equation}
 \pi^0 \rightarrow \gamma + \gamma,
\end{equation}

\noindent whereas the decay of charged pions injects secondary leptons, 

\begin{equation}
 \pi^+ \rightarrow \mu^+ + \nu_{\mu}, \hspace{3mm} \mu^+ \rightarrow \rm{e^+} + \nu_{\rm{e}} + \bar{\nu}_{\mu},\\
\end{equation}
\vspace{-5mm}
\begin{equation}
  \pi^- \rightarrow \mu^- + \bar{\nu}_{\mu}, \hspace{3mm} \mu^- \rightarrow \rm{e^-} + \bar{\nu}_{\rm{e}} + \nu_{\mu}.
\end{equation}

\noindent Secondary pairs are also injected by direct photopair production

\begin{equation}
 \rm{p} + \gamma \rightarrow \rm{p} + \rm{e^-} + \rm{e^+}.
\end{equation}

\noindent Photomeson production against stellar photons is not considered here since the energies of the particles do not reach the threshold. Although stellar photons can have energies greater than the threshold for photopair production, this process does not contribute significantly to the radiative output of the source and thus is also disregarded. 

There is a third source of secondary leptons, namely the annihilation of two photons into an electron-positron pair. This process is also a photon sink for gamma rays, that annihilate with the low energy photons from the star and the jet itself. The effects of absorption on the SED are discussed in more detail in Section \ref{subsec:seds}.

The density of cold protons in the stellar wind is required to calculate the $pp$ cooling rate and the $\pi^0$ emissivity. This is fixed by the continuity equation

\begin{equation}
 \dot{M}_\bigstar = 4\pi r^2 \rho(r)\; v(r),
\end{equation}

\noindent where $r$ is the distance to the center of the star and $\rho$ and $v$ are the mass density and velocity of the wind, respectively. Introducing the standard velocity profile of massive stars \citep{lamers1990}, the proton number density as a function of $z$ results

\begin{equation}
 n(z) = \frac{\dot{M}_\bigstar}{4 \pi\; (a_\bigstar^2 + z^2)\, v_\infty\, m_{\rm{p}}}  \left( 1 - \frac{R_\bigstar}{\sqrt{a_\bigstar^2 + z^2}}\right)^{-1},
\end{equation}

\noindent where $m_p$ is the proton mass, $v_\infty$ is the terminal velocity of the wind,  and $R_\bigstar$ and $\dot{M}_\bigstar$ are the radius  and  the mass-loss rate of the star, respectively. We assume that only a fraction $\chi$ of the matter in the stellar wind is able to mix with the jet. The value of $\chi$ is approximated to be equal to the ratio between the mass density in the jet and in the wind. We obtain $\chi\approx 0.1$. 
  
\subsection{Relativistic particle distributions}

The injection of primary electrons and protons in the jet reference frame is parametrized as a power-law in energy times an exponential cutoff,

\begin{equation}
 Q(E,z) = Q_0\,  E^{-\Gamma}\, \exp{\left[-E/E_{\rm{max}}(z)\right]}.
\end{equation}

\noindent Here, $Q_0$ is a normalization constant obtained from the total power injected in each particle species. The spectral index for diffusive shock acceleration is in the range $1.5\leq \Gamma \leq 2.4$ \citep[see for example][]{rieger2007}. The injection function is different from zero only in the region $z_{\rm acc}\leq z \leq z_{\rm max}$ and in the energy interval $E_{\rm{min}}\leq E \leq E_{\rm{max}}$. The values of  $z_{\rm max}$ and $E_{\rm{min}}$ are free parameters, whereas the maximum energy $E_{\rm{max}}$ is calculated by equating the total energy-loss rate with the acceleration rate \citep[e.g.][]{aharonian2004}

\begin{equation}
 \left.\frac{dE}{dt}\right|_{\rm{acc}} (E, z) = \eta e c B(z),
\end{equation}

\noindent where $e$ is the electron charge, $\eta<1$ accounts for the efficiency of the acceleration mechanism, and $B(z)$ is the magnetic field strength at $z$, calculated as $B(z) = B_0 \left( z_0 / z\right)^m$. 

As discussed in Section \ref{sec:rates}, the interaction of relativistic protons with matter and radiation injects secondary particles into the jet. The injection function of these particles depends on the specific mechanism; see the references in \cite{vila-etal2012} for the complete set of expressions. 

Once $Q(E,z)$ is known, we compute the isotropic, steady-state particle distributions $N(E,z)$ in the jet reference frame for each particle species solving the transport equation \citep{khangulyan2008}

\begin{equation}
 v_{\rm{conv}} \frac{\partial N}{\partial z} + \frac{\partial}{\partial E} \left( \left.\frac{dE}{dt}\right|_{\rm{tot}} N \right) + \frac{N}{\tau_{\rm{dec}}(E)} = Q(E,z).
 \label{eq:transport}
\end{equation}

\noindent The first and second terms on the left-hand side account for the changes in the particle distribution due to convection along the jet and energy losses, respectively. The convection velocity is of order of the jet bulk velocity, $v_{\rm{conv}} \approx v_{\rm{jet}}$.\footnote{Since the jet is very collimated, we neglect the radial component of its bulk velocity compared to the $\hat{z}$-component.} The third term is non-zero only for decaying particles (i.e., pions and muons); $\tau_{\rm{dec}}$ is the mean life-time in the jet frame.  

\subsection{Spectral energy distributions}
\label{subsec:seds}
  
  %
  %
	%
  %
%
  %
  %
  %
  %
  %
  
From the particle distributions, we calculate the radiative output of all species of primary and secondary particles in the jet. Detailed formulae can
be found in \cite{romero-vila2008}, \cite{vila-etal2012} and references therein. Hereafter we use primed and unprimed symbols for quantities measured in 
the jet comoving reference frame and in frame of the observer, respectively. 

For each radiative process (except proton-proton interactions, for which calculations are carried out in the observer frame) we compute the volume
emissivity $q'_\gamma$ (in units of erg s$^{-1}$ cm$^{-3}$ erg$^{-1}$ sr$^{-1}$) in the jet frame, and transform it to the observer frame according 
to $q_\gamma = D^2 q'_\gamma$ \citep[e.g.][]{RybickiLightman1986, Lind1985}. Here

\begin{equation}   
D = \left[ \Gamma_{\rm{jet}} \left( 1 - \beta_{\rm{jet}} \cos \theta_{\rm{jet}}\right) \right]^{-1}
\end{equation}
  
\noindent is the Doppler factor, $\theta_{\rm{jet}}$ is the jet viewing angle and \mbox{$\beta_{\rm{jet}} =  v_{\rm{jet}}/c$}. We then obtain the 
specific luminosity (in $\rm{erg~s^{-1}~sr^{-1}}$) at photon energy $E_\gamma$ as

\begin{equation}
L_\gamma (E_\gamma) = E_\gamma\, \int_{V} q_\gamma\, dV,  
\end{equation}

\noindent where $V$ is the volume of the emission region and the photon energy in the observer frame is related that in the comoving frame as
  
\begin{equation} 
E_\gamma = D E'_\gamma.  
\end{equation}
	
  This intrinsic luminosity must be corrected for absorption caused by photon-photon annihilation into pairs, \mbox{$\gamma+\gamma \rightarrow e^+ + e^-$},  that affects mainly the gamma-ray band of the SED. The target low-energy photons are those of the internal radiation field of the jet (mainly primary electron synchrotron emission), the accretion disk and the companion star.
The calculation of the optical depth intrajet is carried out in the local approximation of \citet{Ghisellini85} for the density of the target fields; for external absorption we follow \cite{becker1995}. As already mentioned we assume that the system is in the superior conjunction. The effect of the orbital motion of the binary is discussed below. 

\begin{table*}[htbp]
\caption{Values of the model parameters.}
\label{tab:parameters}
\centering
\begin{tabular}{l c c}
\hline
Parameter & Symbol & Value \\
\hline
\multicolumn{3}{c}{Binary parameters}\\
\hline
Black hole mass & $M_{\rm{BH}}$ & $14.8~M_{\odot}$\\
Binary separation & $a_\bigstar $ &$3.2 \times 10^{12}$ cm\\
Star mass & $M_\bigstar$ & $20~M_{\odot}$\\
Star radius & $R_\bigstar$ & $19~R_{\odot}$\\
Star temperature & $T_\bigstar$ & $2.8\times 10^4~\rm{K}$\\  
Terminal wind velocity & $v_\infty$ & $2500~\rm{km~s^{-1}}$ \\
Star mass-loss rate& $\dot{M}_\bigstar$& $10^{-5}~M_{\odot}~\rm{yr}^{-1}$\\
Distance to Earth & $d$ & $1.86~\rm{kpc}$\\
\hline
\multicolumn{3}{c}{Jet fixed parameters}\\
\hline 
Jet viewing angle & $\theta_{\rm{jet}}$ & $29^\circ$\\
Jet opening angle & $\theta_{\rm op} $ & $2^\circ$\\
Jet injection radius & $r_0 $ &  $3.3\times 10^6$ cm \\
Base of the jet & $z_0 $ & $1.1\times 10^8~\rm{cm}$ \\
Base of the acceleration region & $z_{\rm{acc}} $ & $2.8\times 10^8~\rm{cm}$\\
End of the jet & $z_{\rm{end}} $ &$1.0 \times 10^{15}~\rm{cm}$\\
Jet bulk Lorentz factor & $\Gamma_{\rm jet}$ & $1.25$\\
Magnetic field jet base & $B_0$ & $5\times 10^7$ G \\
Jet power  & $L_{\rm jet}$ & $4\times 10^{37} - 10^{38}$ erg s$^{-1}$ \\
Ratio $L_{\rm rel}/L_{\rm jet}$ & $q_{\rm rel}$ & 0.1\\
Jet-wind entrainment factor & $\chi$ & $0.1$ \\
\hline
\multicolumn{3}{c}{Jet free parameters}\\
\hline
Magnetic field decay index & $m$ & $1-2$\\
Particle injection spectral index & $\Gamma$ & $1.5-2.2$\\
Acceleration efficiency & $\eta$ & $10^{-4}-10^{-1}$\\
Minimum particle energy [$mc^2$] & $E_{\rm{min}}$ & $2 - 120$\\
Ratio $L_p/L_e$ & $a$ & $10^{-2} - 10^2$\\
End of acceleration region & $z_{\rm{max}} $ &$10^{10} - 10^{14}$ cm\\
\hline
\end{tabular}
\end{table*}

\section{Results}
\label{sec:results}

We performed least-squares fits to the observational data by varying some of the model parameters ($z_{\rm max}$, $m$, $a$, $\Gamma$, $\eta$ and $E_{\rm min}$)
in the intervals shown in Table \ref{tab:parameters}. The rest of the parameters were kept fixed. The value of $z_{\rm{end}} = 1.0 \times 10^{15}~\rm{cm}$, in particular, was chosen in order to match the extension of the jet with the size inferred from the radio images by \cite{stirling2001}. The fitting algorithm searches for the minimum of the sum of the quadratic distances between each data point and the value of the luminosity predicted by the model at the same energy. Upper limits in the gamma-ray domain are also included in the fit: models that predict gamma-ray luminosities above the observed upper limits are strongly penalized by assigning them a large value of the figure of merit. Observational errors were not considered since they were not available for all the data points. 

\begin{figure*}[htp]
 \centering
 \includegraphics[width = 0.9\textwidth]{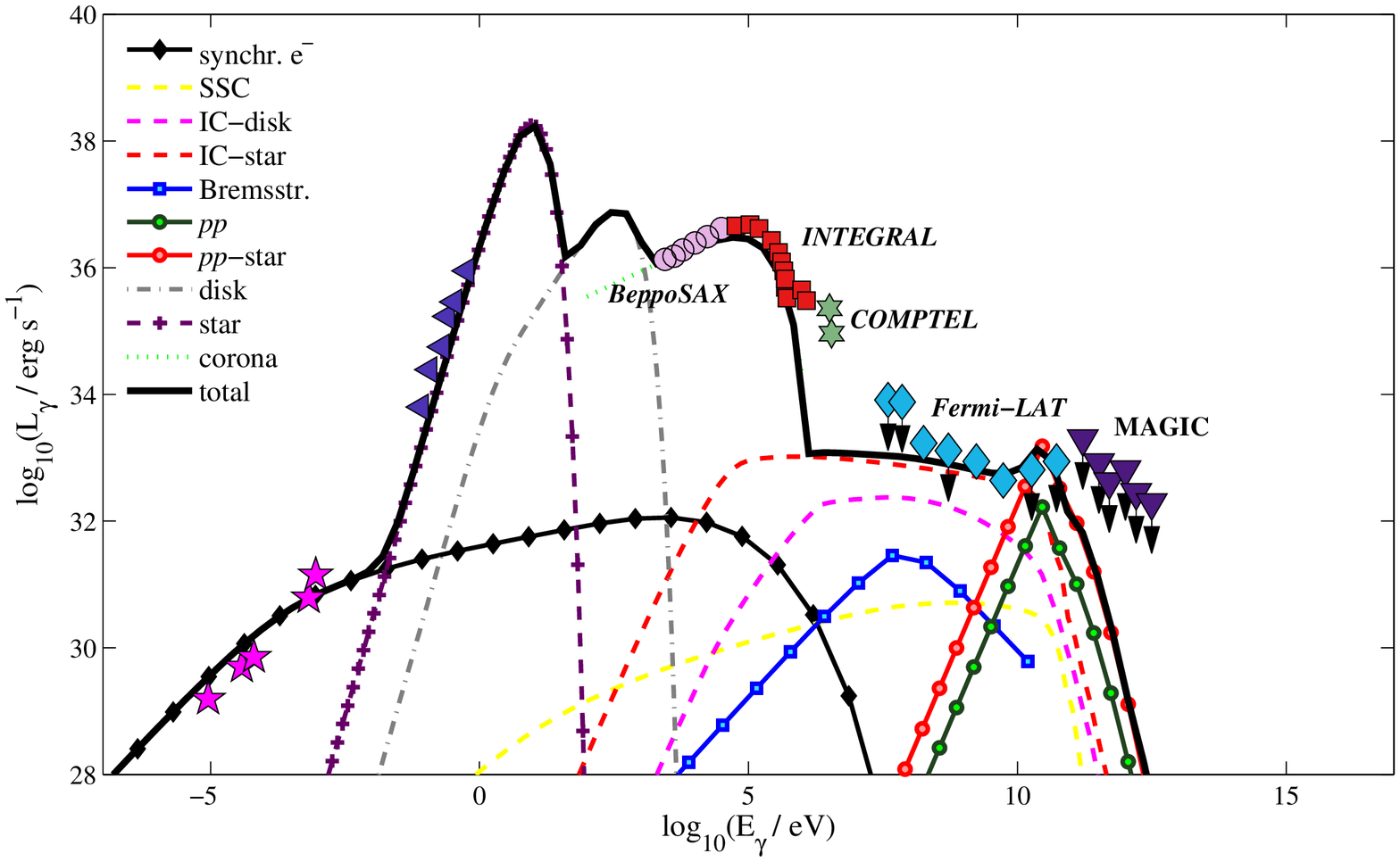}
 \includegraphics[width = 0.9\textwidth]{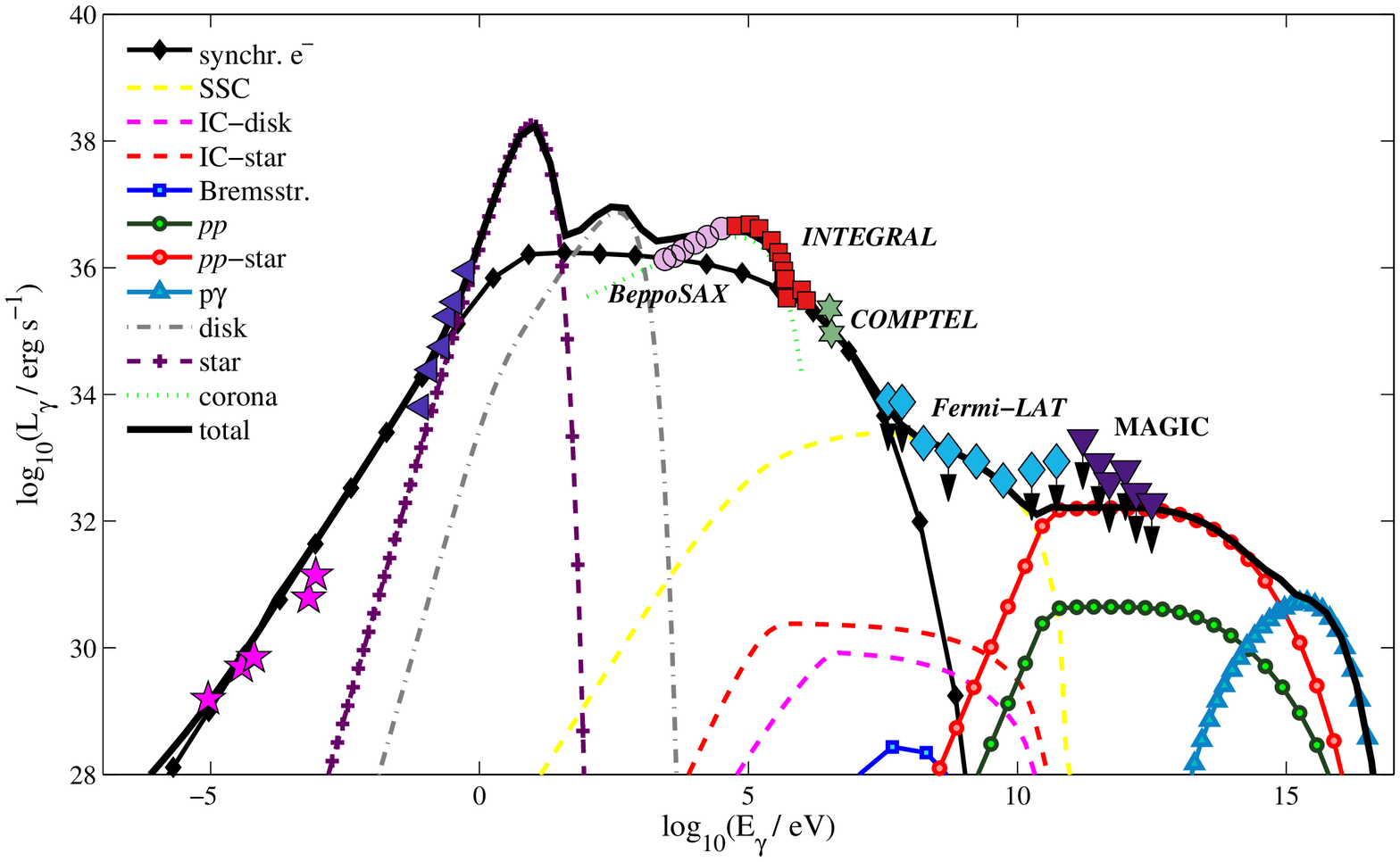}
 \caption{Best-fit spectral energy distributions for Cygnus X-1. The top panel corresponds to Model A and the bottom panel to Model B. Radio data is taken from \cite{fender2000}, IR fluxes from \cite{persi1980} and \cite{mirabel1996}, 
soft X-ray observations below $5 \times 10^{18}$ Hz (by \textit{BeppoSAX}) from \cite{diSalvo2001}, hard X-ray data above $20$~keV (by \textit{INTEGRAL}) from \cite{zdziarski2012}, soft gamma-ray data  (by \textit{COMPTEL}) from \cite{mcConnell2002}, the \textit{Fermi}-LAT measurements and upper limits from \cite{malyshev2013}, and the MAGIC upper limits from \cite{albert2007}. Down-pointing arrows indicate upper limits. The data are not simultaneous.}
 \label{fig:SEDs}
\end{figure*}

Two different sets of parameters led to equally good fits; a detailed list of the values of the best fit parameters is given in Table \ref{tab:models}. The corresponding SEDs are shown in Fig. \ref{fig:SEDs} along with the observational data; the characteristics of the SEDs are discussed below in Setion \ref{sect:sed}. The first set (Model A) corresponds to a soft injection of 
relativistic particles in combination with a fast-decaying magnetic field. The particles are accelerated in an extended region and the jet content is highly hadronic. The second set (Model B) corresponds to a harder particle injection spectral index and a low magnetic field decay index. The particles are accelerated more efficiently and the acceleration region results narrower than in Model A. In this model the jet has a low hadronic content.

\begin{table*}
 \caption{Best-fit parameters for Models A and B.}
\label{tab:models}
\centering
\begin{tabular}{l c c c}
\hline
Parameter & Symbol & Model A & Model B \\
\hline
Particle injection spectral index & $\Gamma$ & $2.4$ & $2.0$\\
Magnetic field decay index & $m$ & $1.9$ & $1.0$\\
Acceleration efficiency & $\eta$ & $6\times10 ^{-4}$ & $3\times10 ^{-3}$\\
Minimum particle energy [$mc^2$] & $E_{\rm{min}}$ & $95.4$ & $120$\\
Ratio $L_p/L_e$ & $a$ & $39$ & $0.07$\\
End of acceleration region & $z_{\rm{max}} $ &$1.9 \times 10^{12}~\rm{cm}$ & $8.7 \times 10^{11}~\rm{cm}$\\
\hline
\end{tabular}
\end{table*}

\subsection{Cooling rates}

\begin{figure*}[htp]
 \centering
 \includegraphics[width = 0.33\textwidth, trim=0 0 0 0, clip]{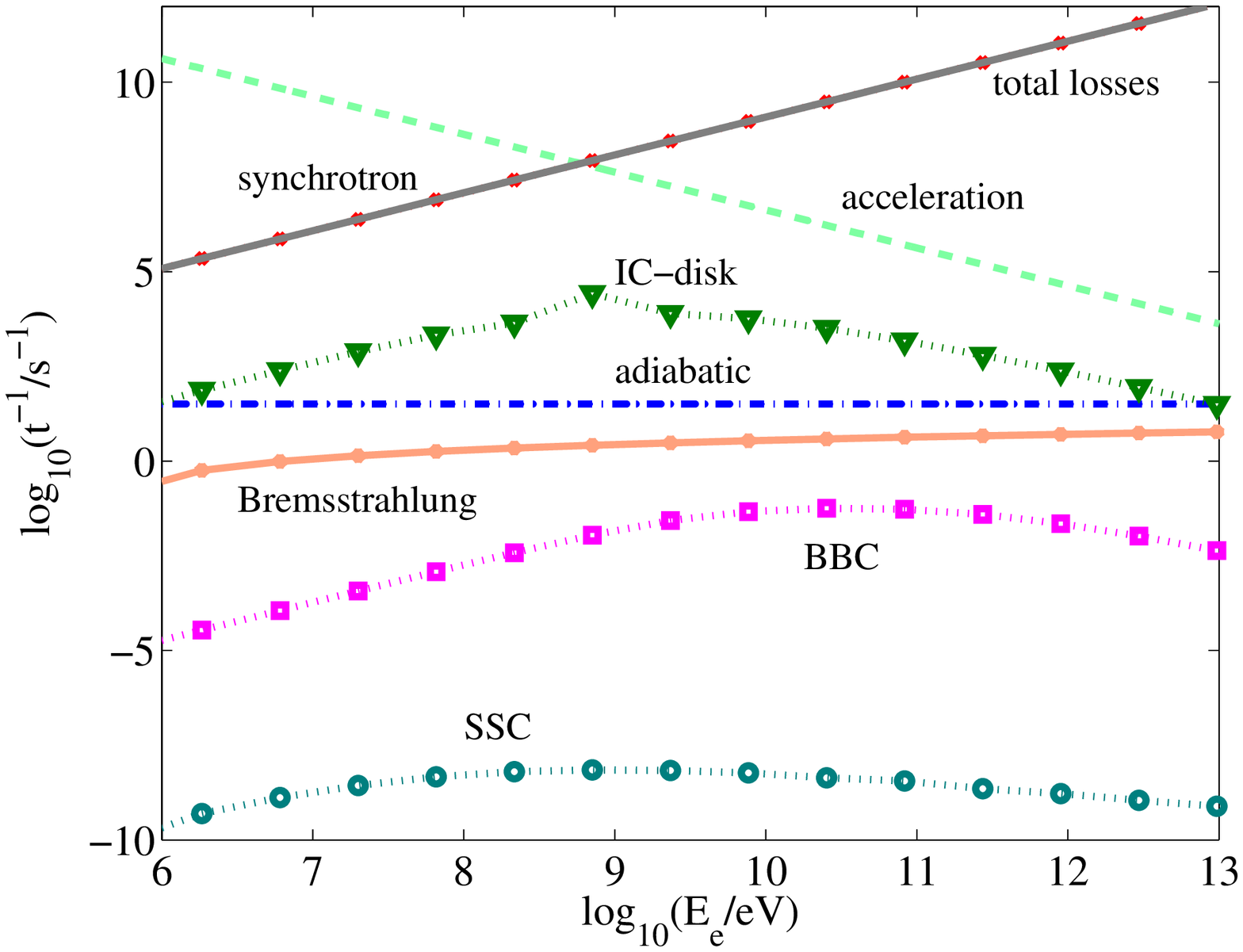}
 \includegraphics[width = 0.33\textwidth, trim=0 0 0 0, clip]{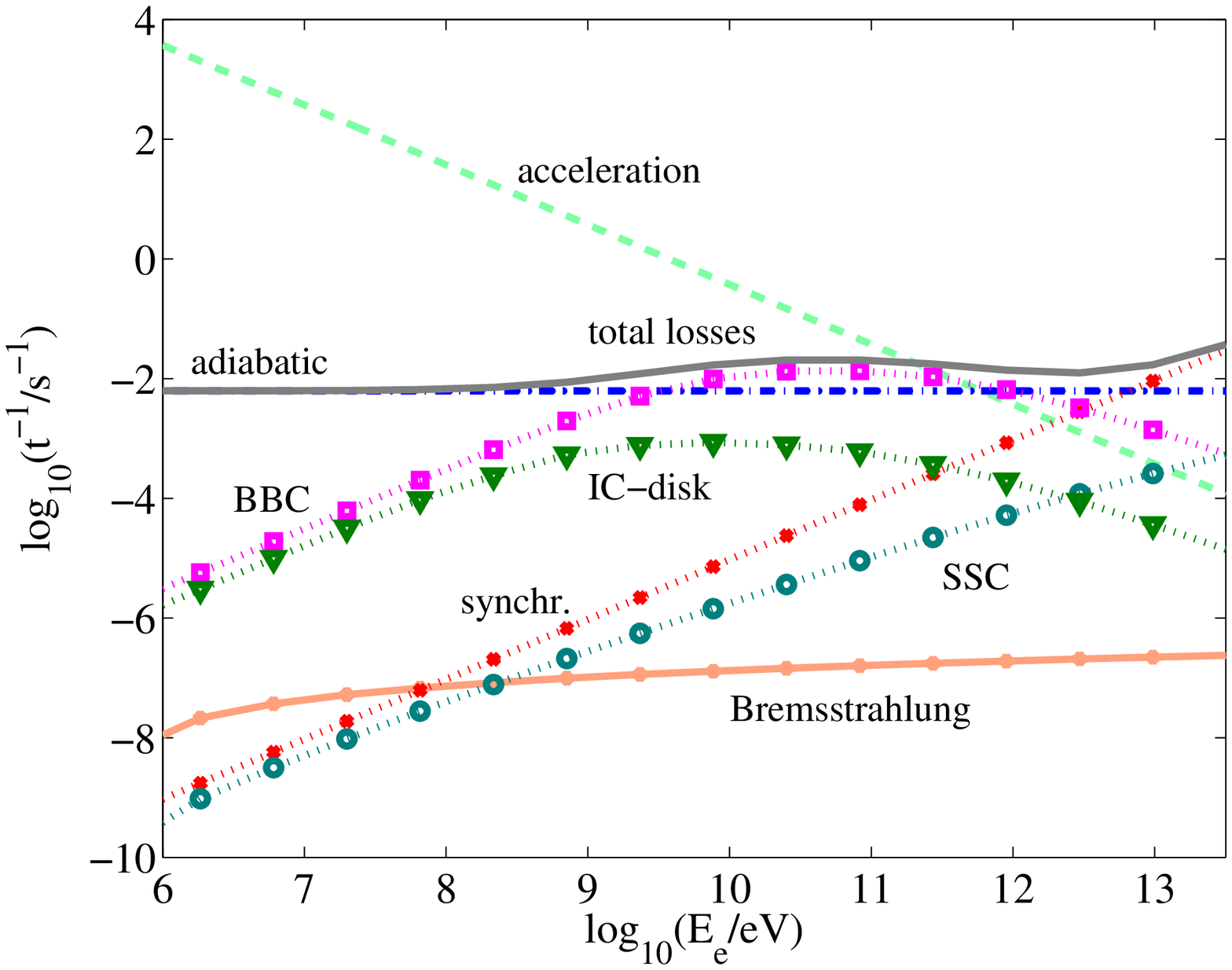}
 \includegraphics[width = 0.33\textwidth, trim=0 0 0 0, clip]{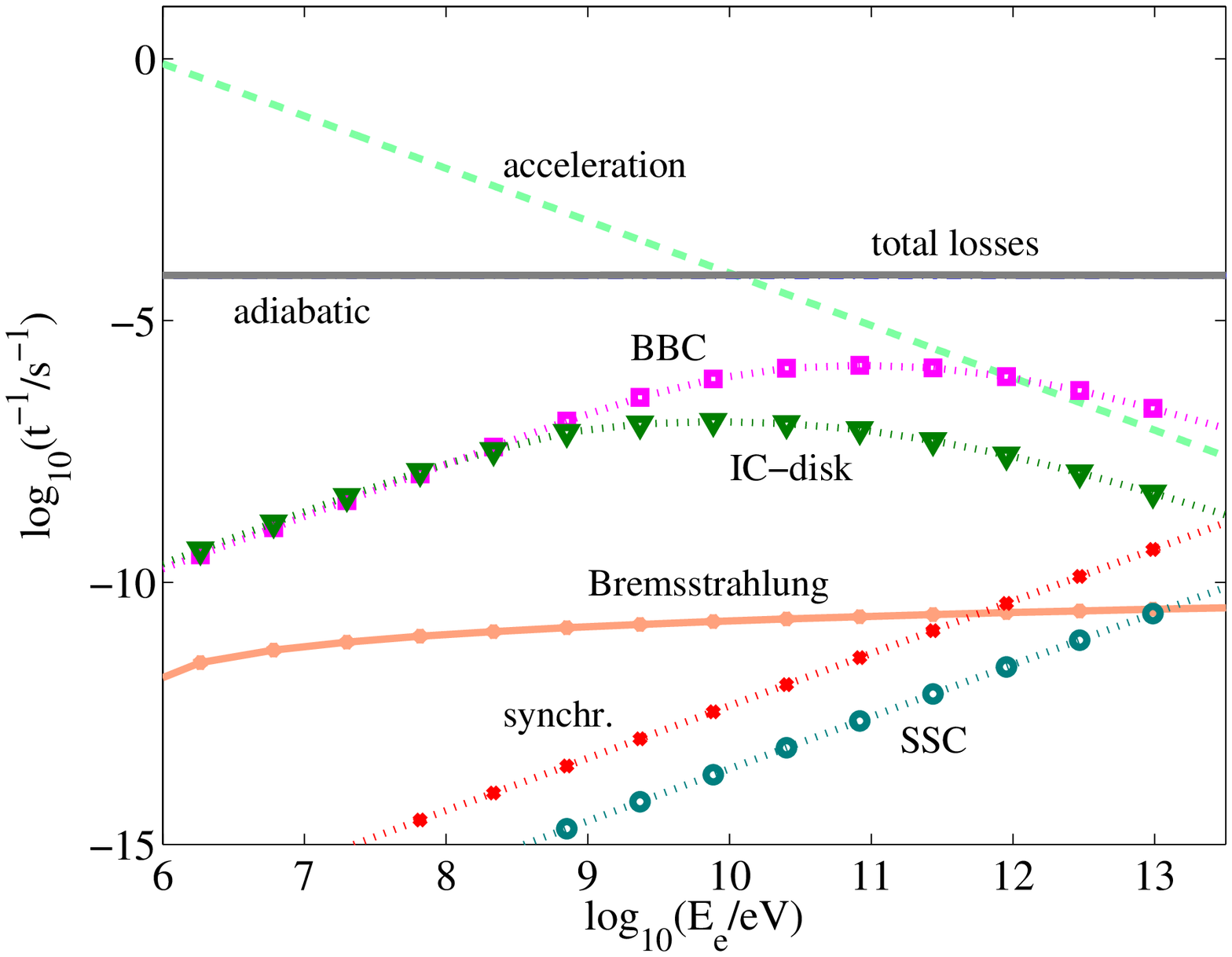}
 \caption{Cooling rates for primary electrons at the base of the acceleration region (left panel), at the end of the acceleration region  (middle panel) and at the end of the jet (right panel) for Model A. }
 \label{fig:rates_electrons_modelA}
\end{figure*}

\begin{figure*}[htp]
 \centering
 \includegraphics[width = 0.33\textwidth, trim=0 0 0 0, clip]{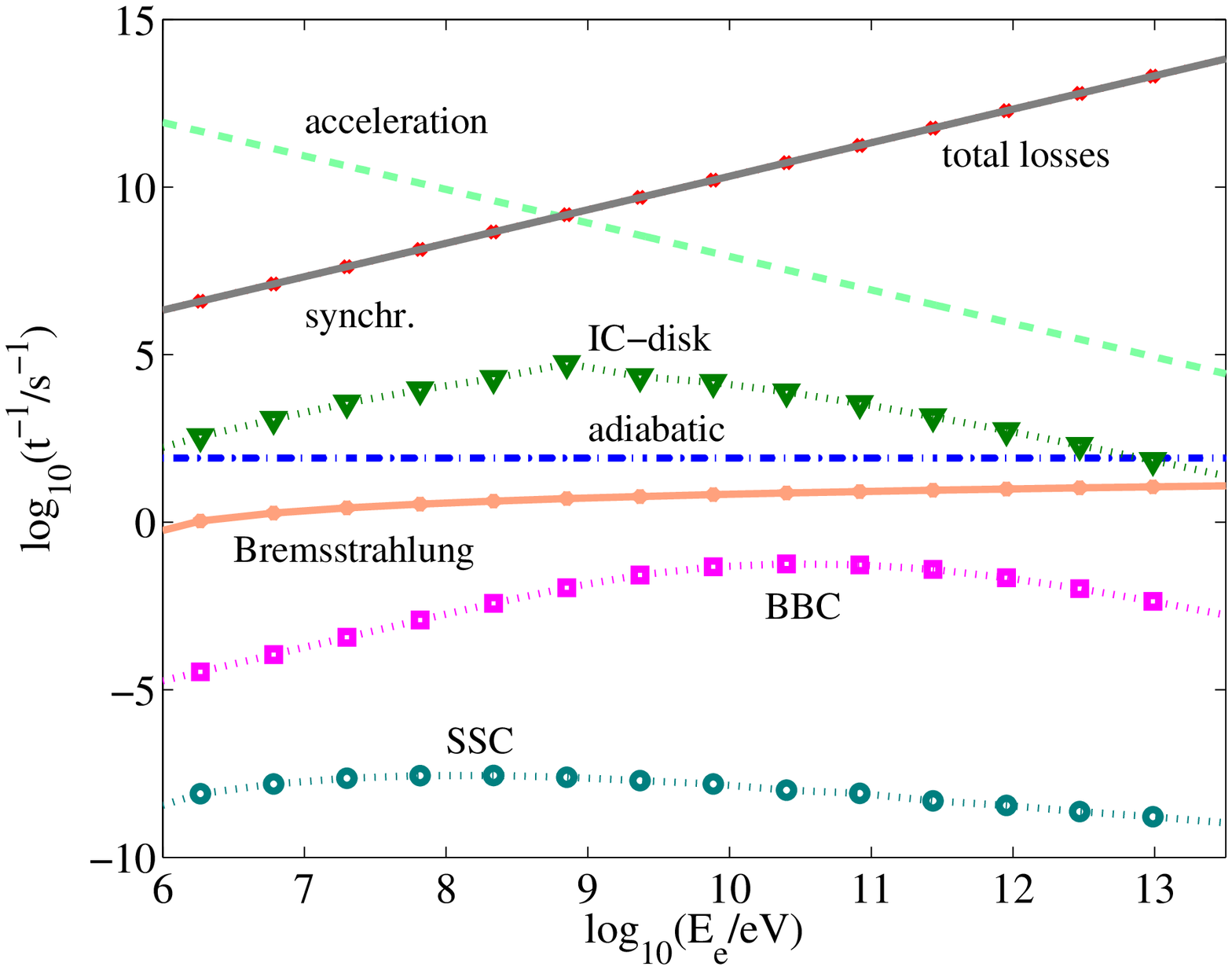}
 \includegraphics[width = 0.33\textwidth, trim=0 0 0 0, clip]{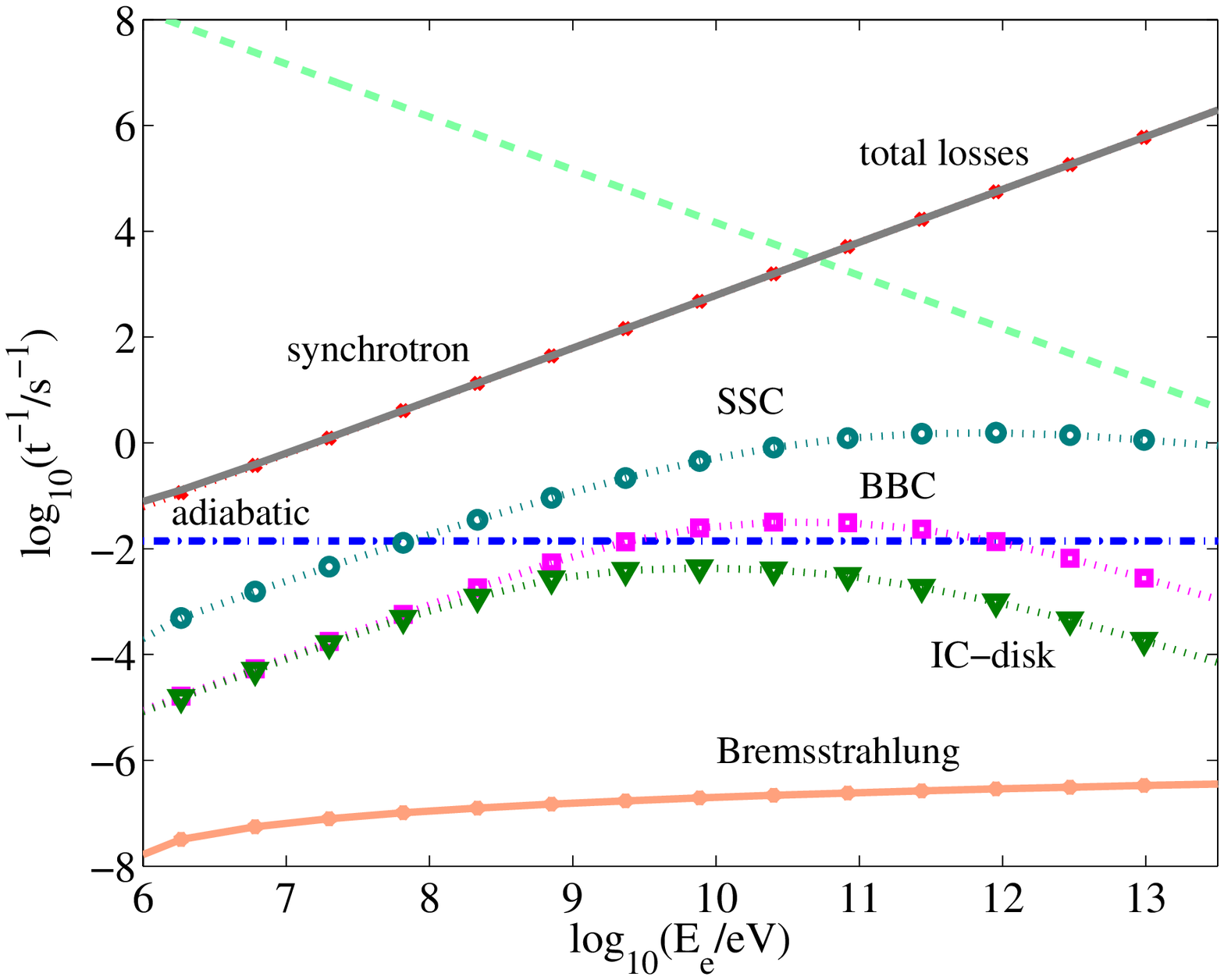}
 \includegraphics[width = 0.33\textwidth, trim=0 0 0 0, clip]{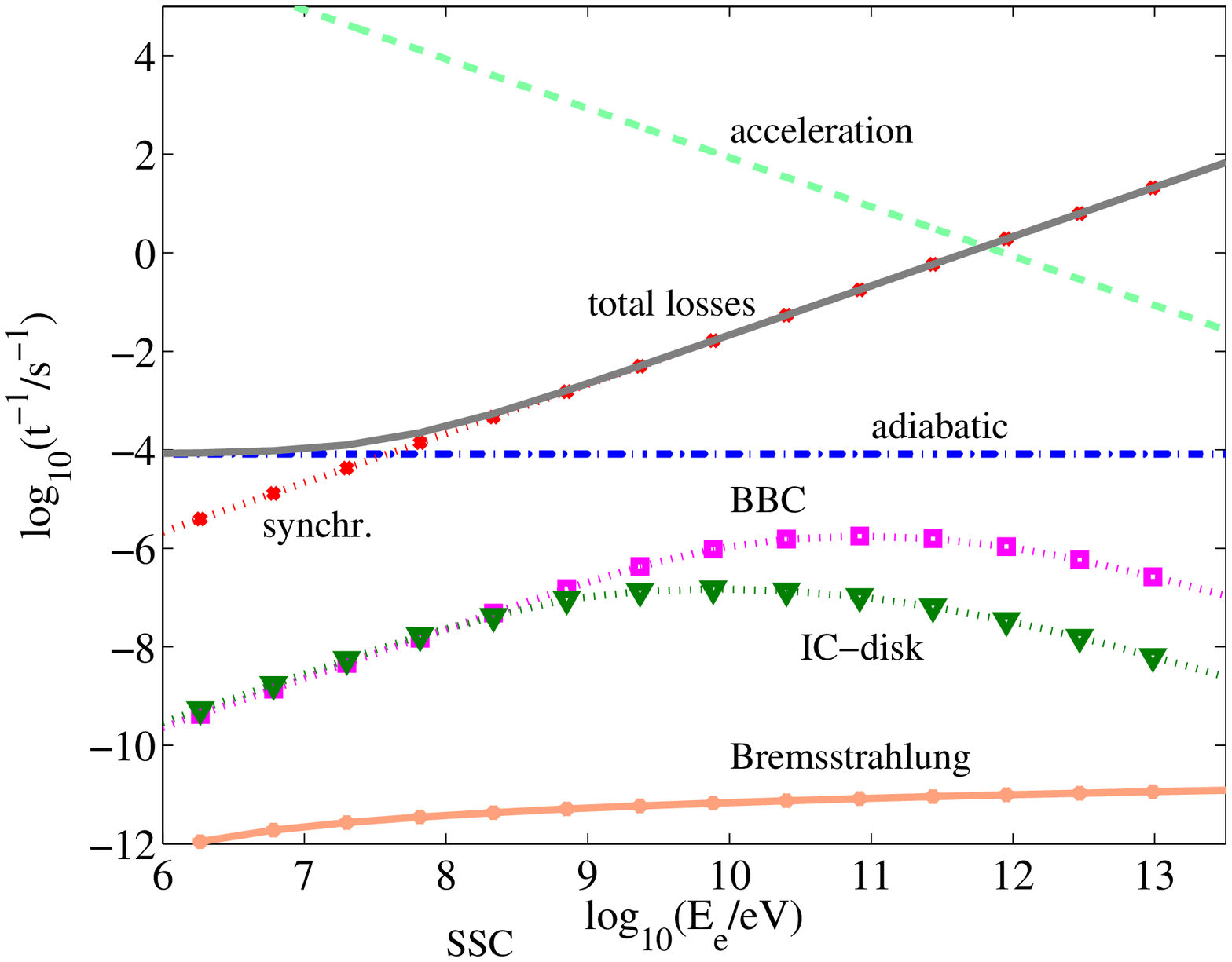}
 \caption{Cooling rates for primary electrons at the base of the acceleration region (left panel), at the end of the acceleration region (middle panel) and at the end of the jet (right panel) for Model B. }
 \label{fig:rates_electrons_modelB}
\end{figure*}

\begin{figure*}[htp]
 \centering
 \includegraphics[width = 0.33\textwidth, trim=0 0 0 0, clip]{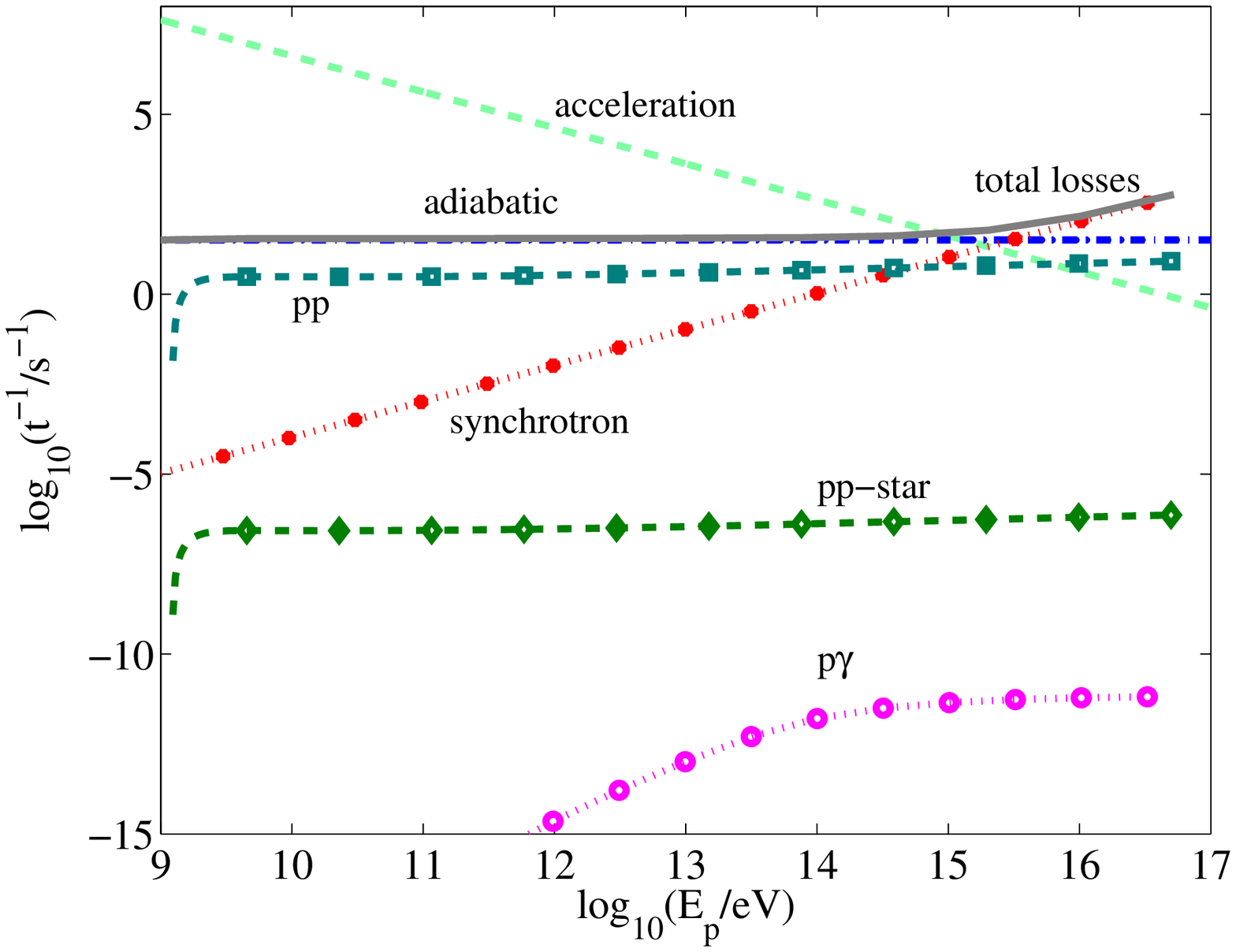}
 \includegraphics[width = 0.33\textwidth, trim=0 0 0 0, clip]{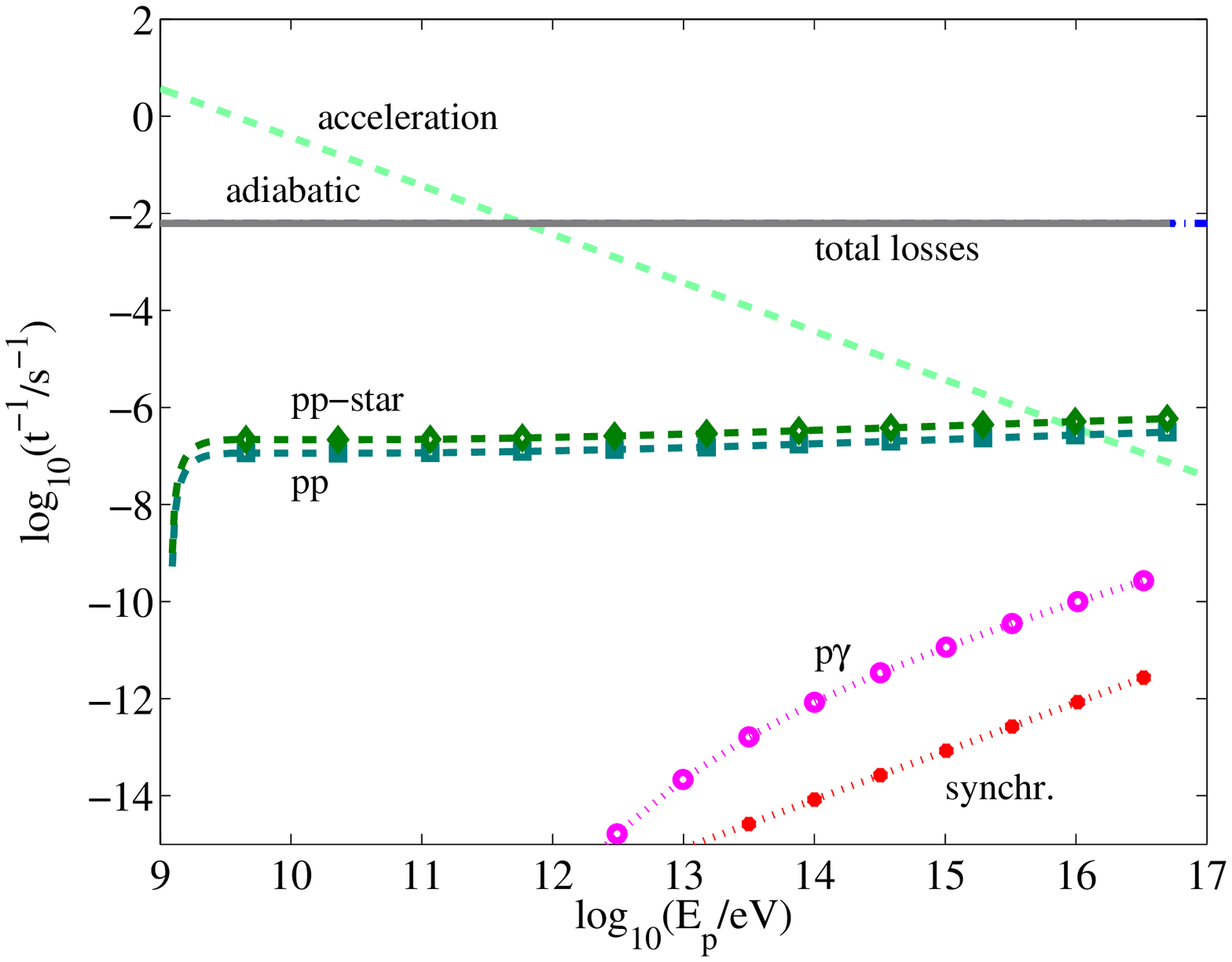}
 \includegraphics[width = 0.33\textwidth, trim=0 0 0 0, clip]{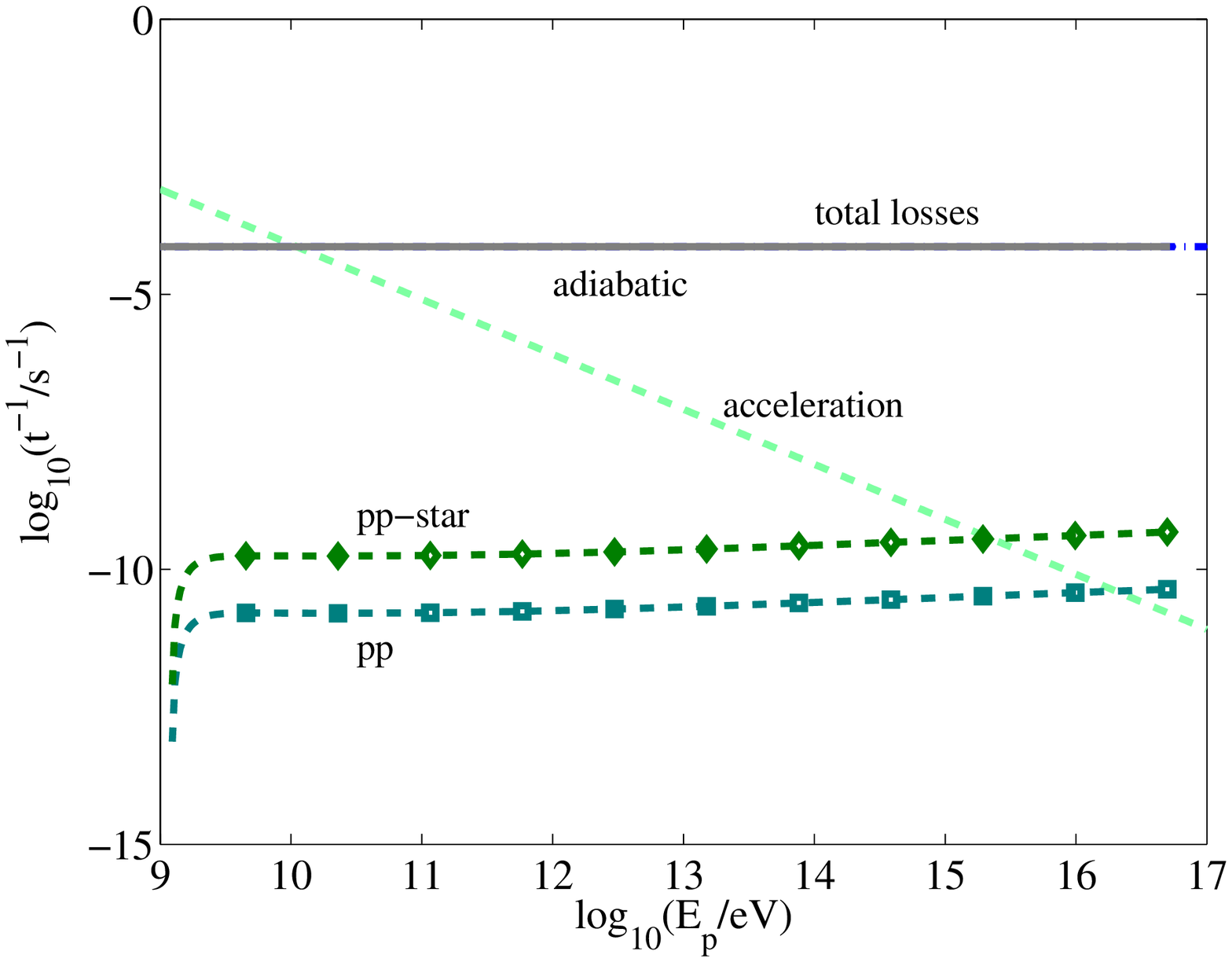}
 \caption{Cooling rates for primary protons at the base of the acceleration region (left panel), at the end of the acceleration region  (middle panel) and at the end of the jet (right panel) for Model A. }
 \label{fig:rates_protons_modelA}
\end{figure*}

\begin{figure*}[htp]
 \centering
 \includegraphics[width = 0.33\textwidth, trim=0 0 0 0, clip]{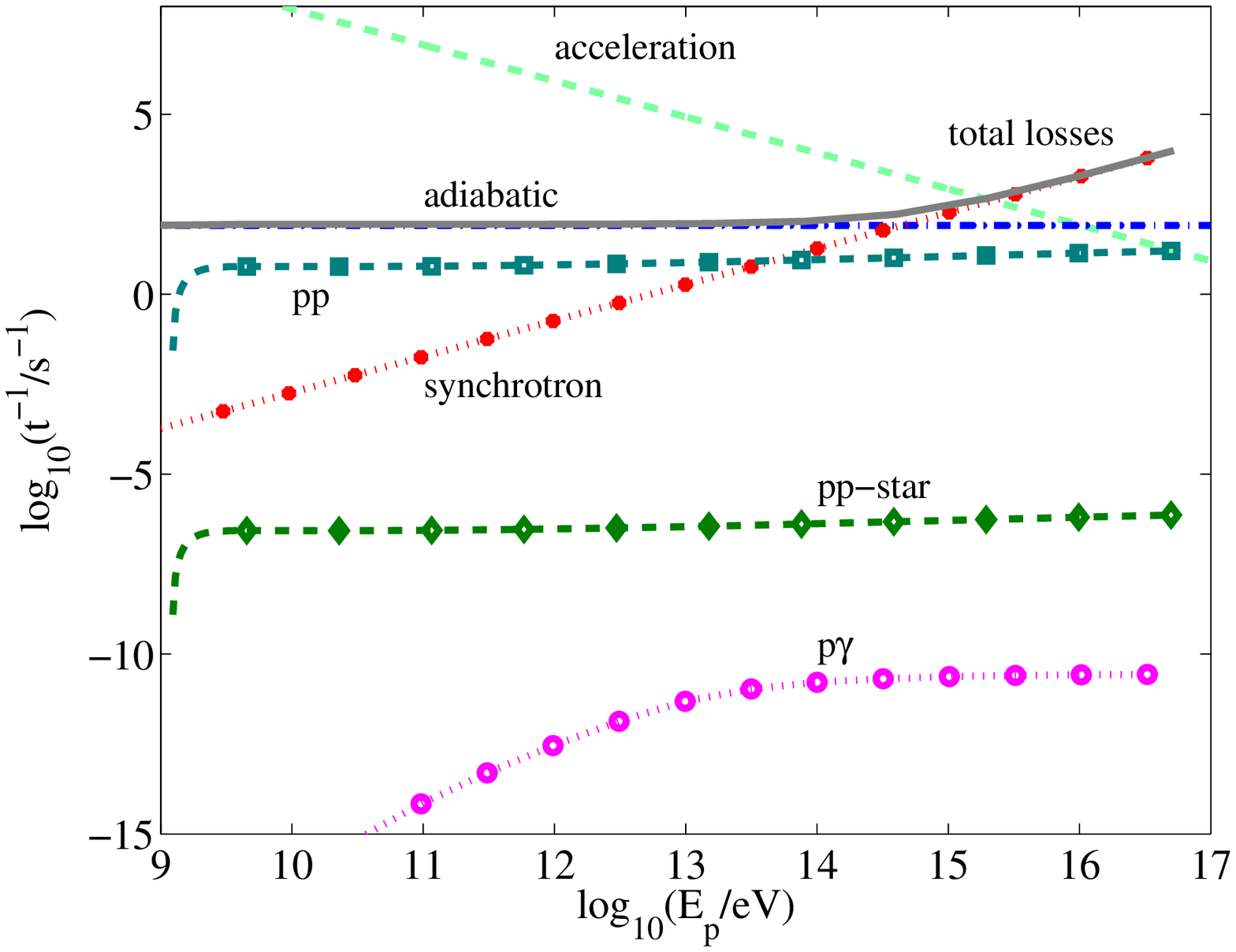}
 \includegraphics[width = 0.33\textwidth, trim=0 0 0 0, clip]{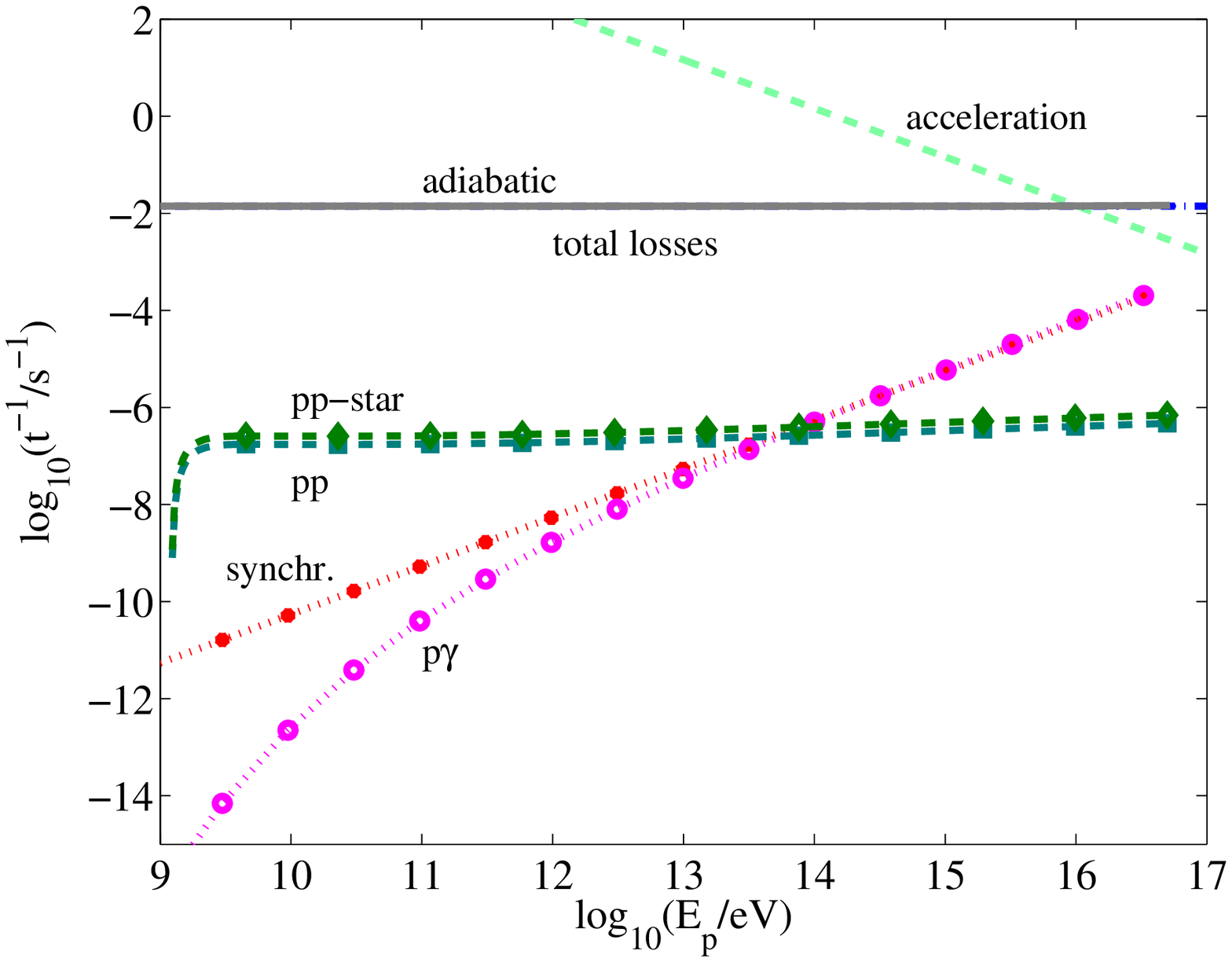}
 \includegraphics[width = 0.33\textwidth, trim=0 0 0 0, clip]{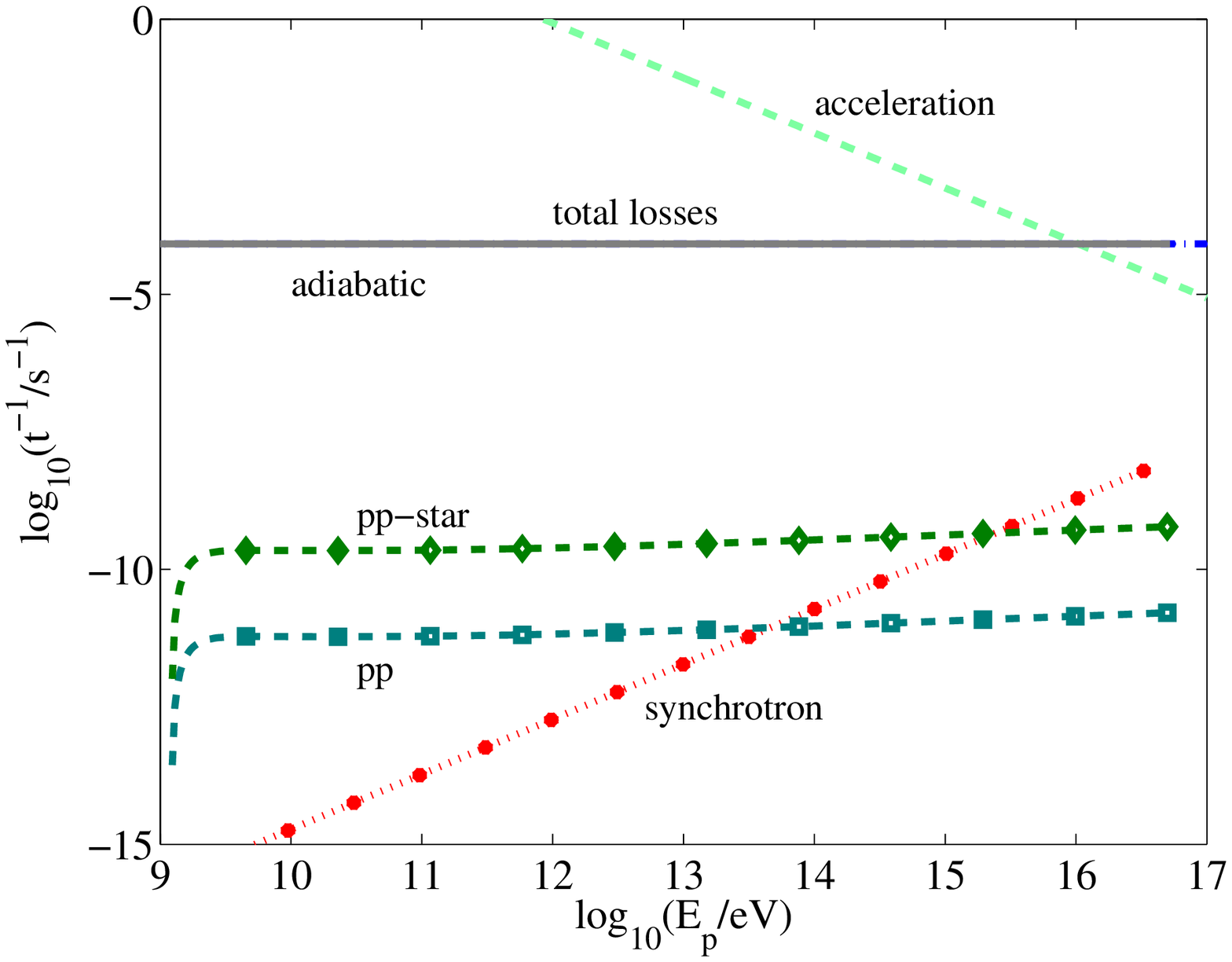}
 \caption{Cooling rates for primary protons at the base of the acceleration region (left panel), at the end of the acceleration region  (middle panel) and at the end of the jet (right panel) for Model B.}
 \label{fig:rates_protons_modelB}
\end{figure*}

In Figs. \ref{fig:rates_electrons_modelA} and \ref{fig:rates_electrons_modelB} we show the cooling rates for primary electrons at different heights in the jet: $z = z_{\rm{acc}}$ (base of the acceleration region), $z = z_{\rm{max}}$ (end of the acceleration region) and $z = z_{\rm{end}}$  (end of the jet). Figures \ref{fig:rates_protons_modelA} and \ref{fig:rates_protons_modelB} show the same for primary protons. At the base of the acceleration region the electron 
cooling is dominated by synchrotron radiation at all energies in both models. In the case of Model A, however, since the magnetic field decays fast, this process loses relevance as the end of the acceleration region is approached, and adiabatic losses take over. 
In the case of Model B, the maximum energy for electrons is always determined by synchrotron losses as a consequence of the slowly decaying magnetic field. Only at the end of the jet and for the lower energies, adiabatic losses become dominant. 
In the case of protons, the cooling is mainly dominated by adiabatic losses, even at the base of the acceleration region. 

\subsection{Particle distributions}

\begin{figure*}[htp]
 \centering
 \includegraphics[width = 0.48\textwidth, trim=0 0 0 0, clip]{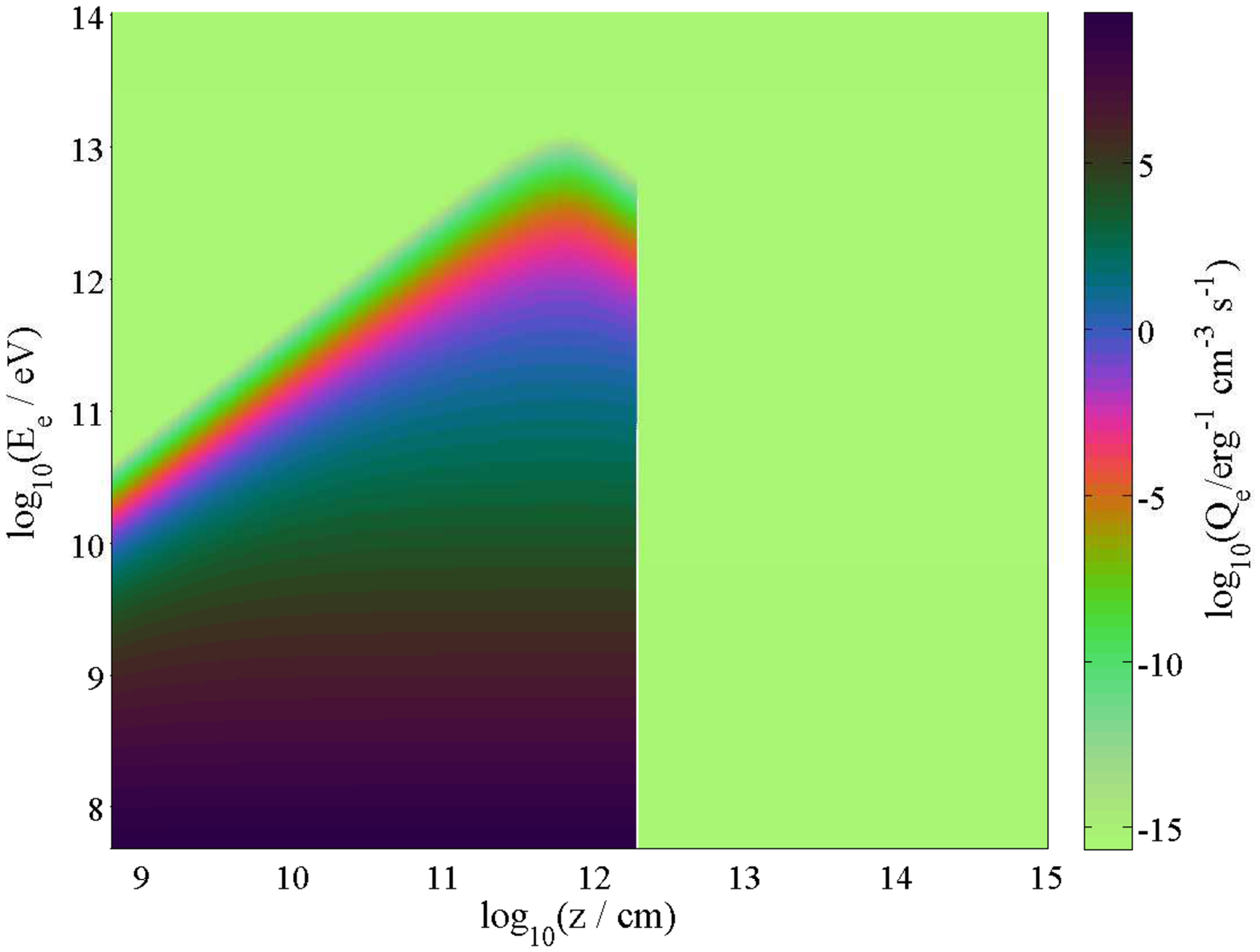}
  \includegraphics[width = 0.48\textwidth, trim=0 0 0 0, clip]{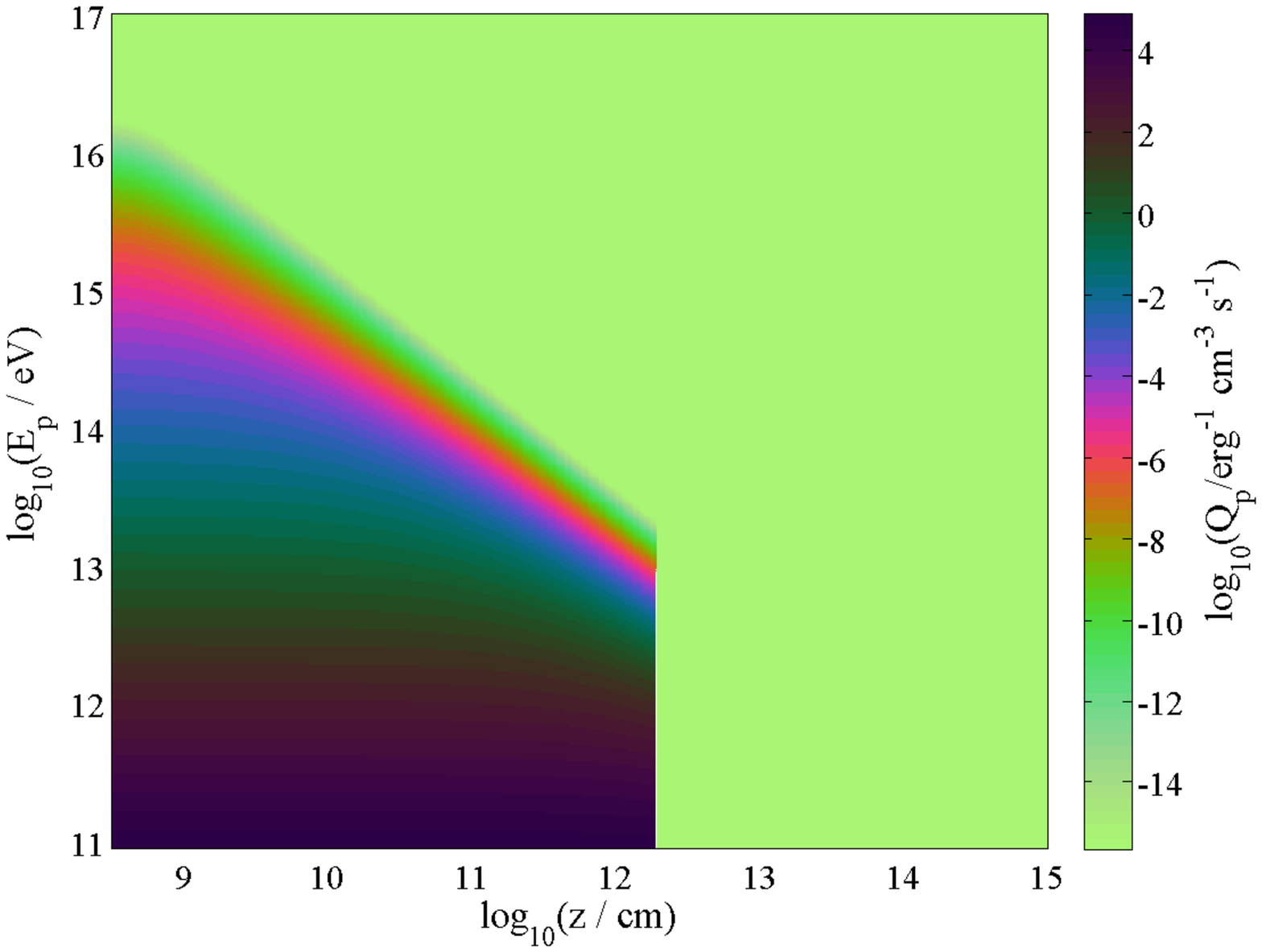}
  \includegraphics[width = 0.48\textwidth, trim=0 0 0 0, clip]{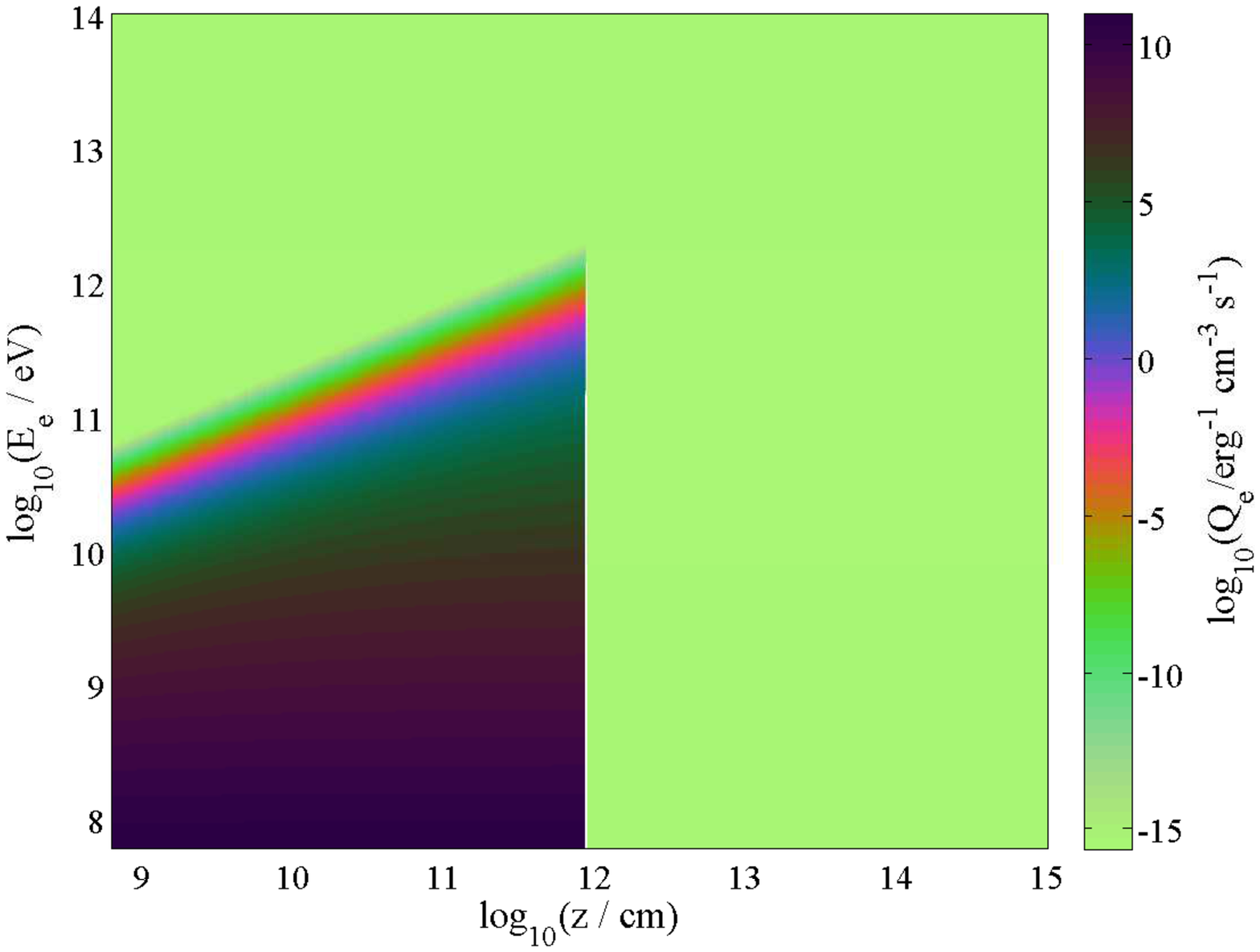}
  \includegraphics[width = 0.48\textwidth, trim=0 0 0 0, clip]{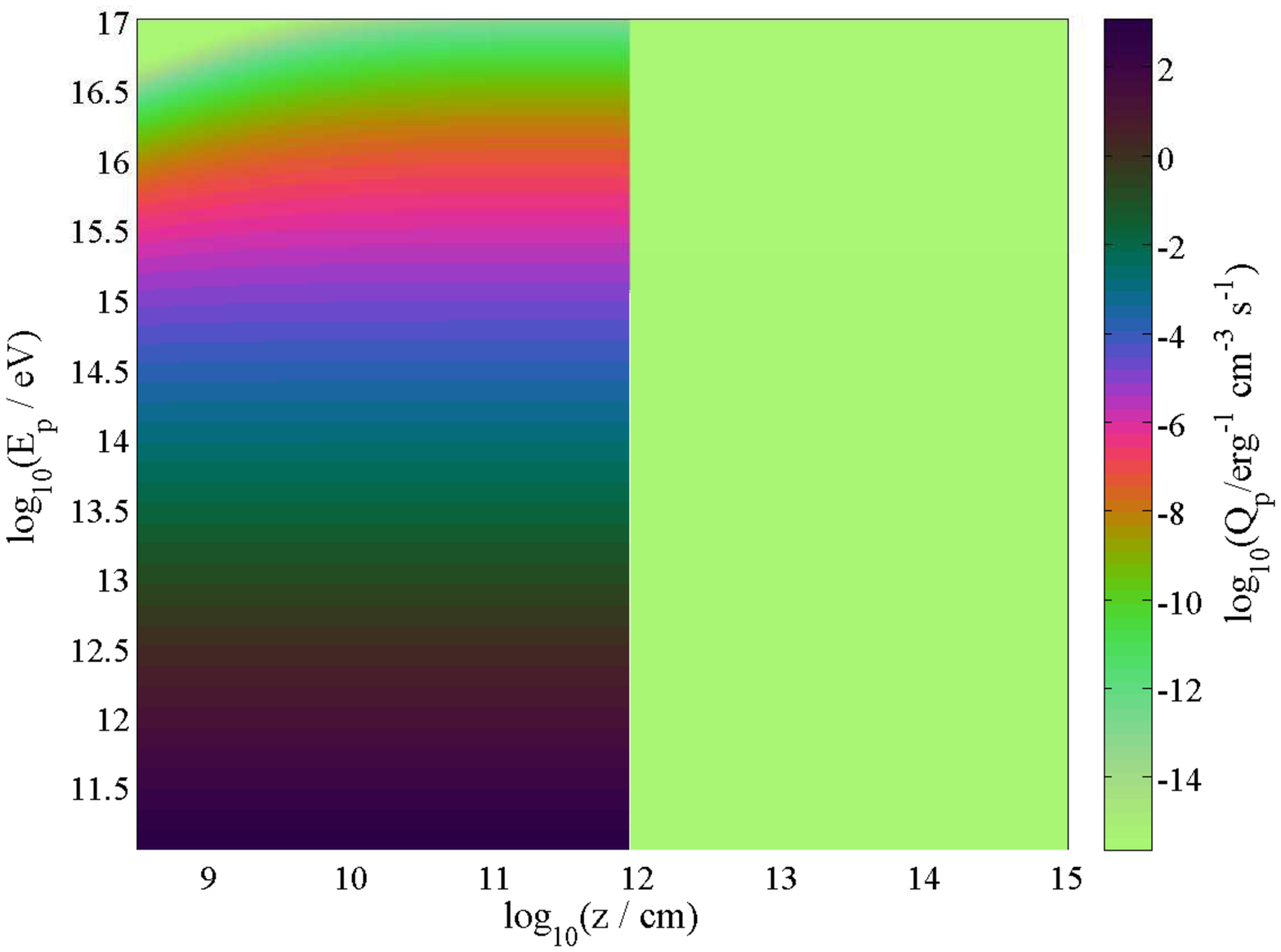}
 \caption{Injection function for primary electrons (left) and protons (right). Top panels correspond to Model A and bottom
panels to Model B. }
 \label{fig:injection}
\end{figure*}

In Fig. \ref{fig:injection} we show the injection function of primary electrons and protons for Models A 
and B. As expected, the maximum energies are higher for protons than for electrons. The maximum electron energy is determined by synchrotron losses and increases with $z$ as the magnetic field decreases,  except near the end of the acceleration region in Model A where adiabatic losses become dominant. Adiabatic losses are the main cooling channel for protons. In Model B the magnetic field decay index is $m=1$, so both the acceleration rate and the adiabatic cooling rate are $\propto z^{-1}$ and the maximum proton energy remains nearly constant all throughout the acceleration region. 

\begin{figure*}[htbp]
 \centering
 \includegraphics[width = 0.48\textwidth, trim=0 0 0 0, clip]{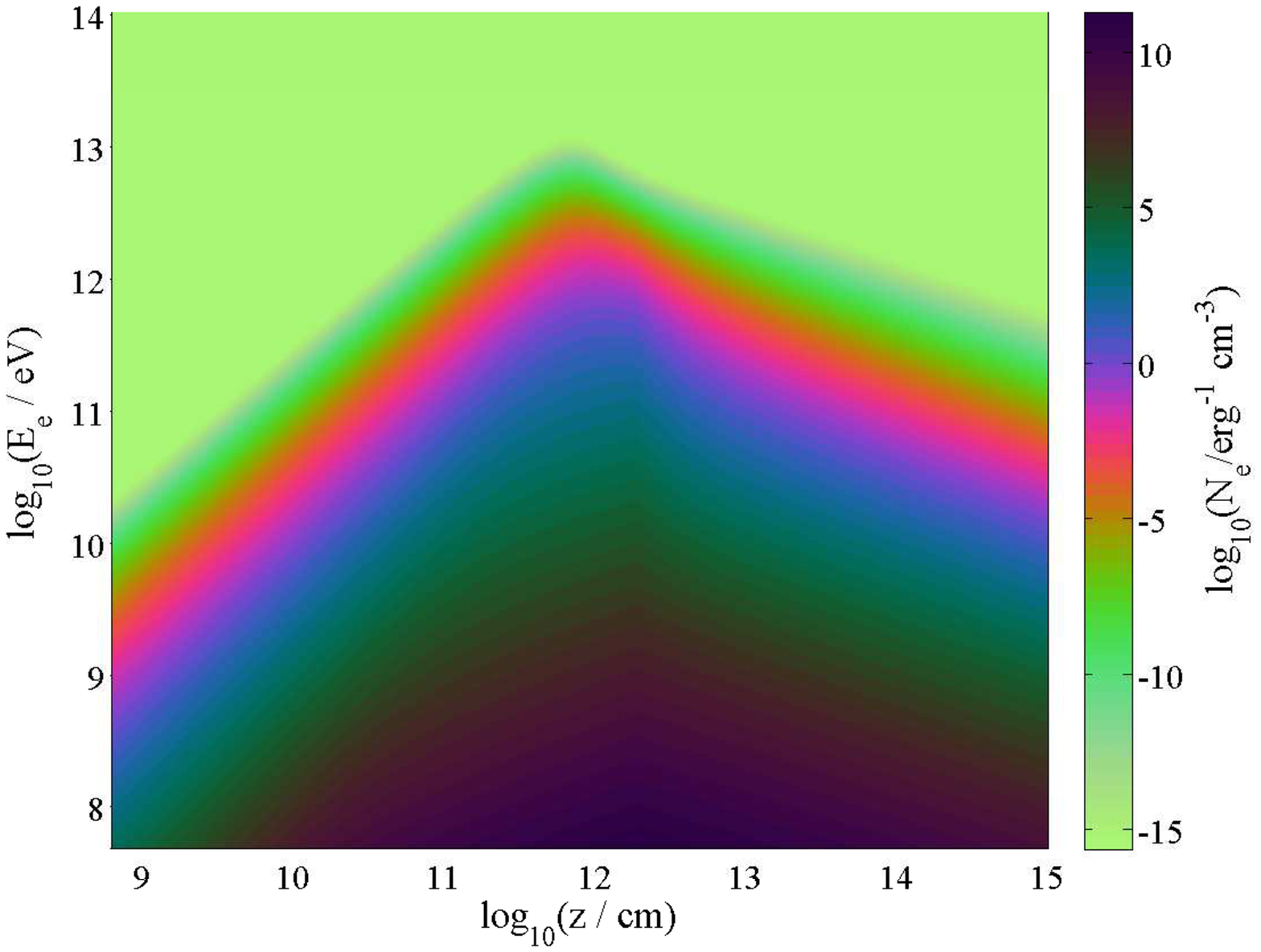}
  \includegraphics[width = 0.48\textwidth, trim=0 0 0 0, clip]{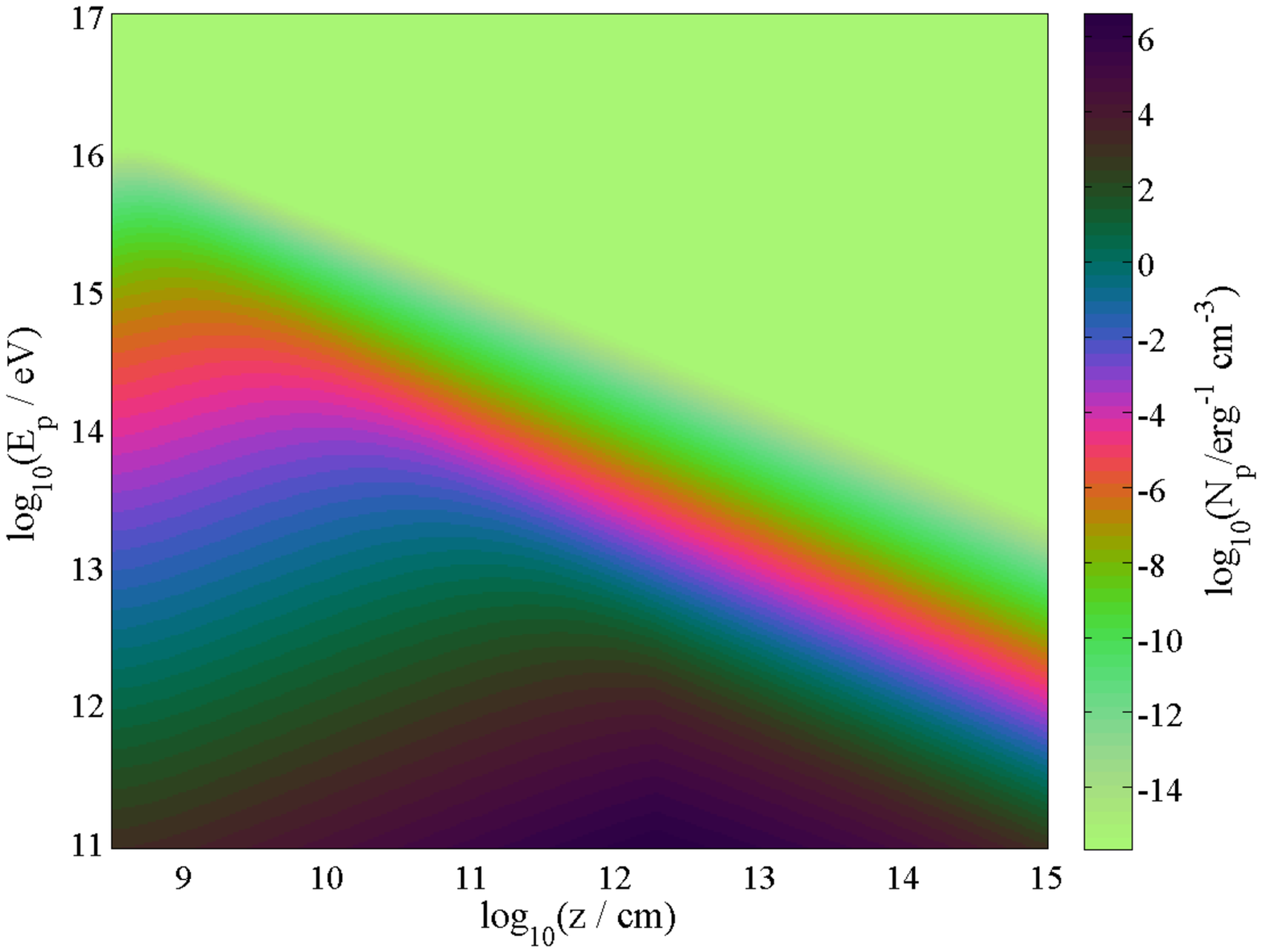}
  \includegraphics[width = 0.48\textwidth, trim=0 0 0 0, clip]{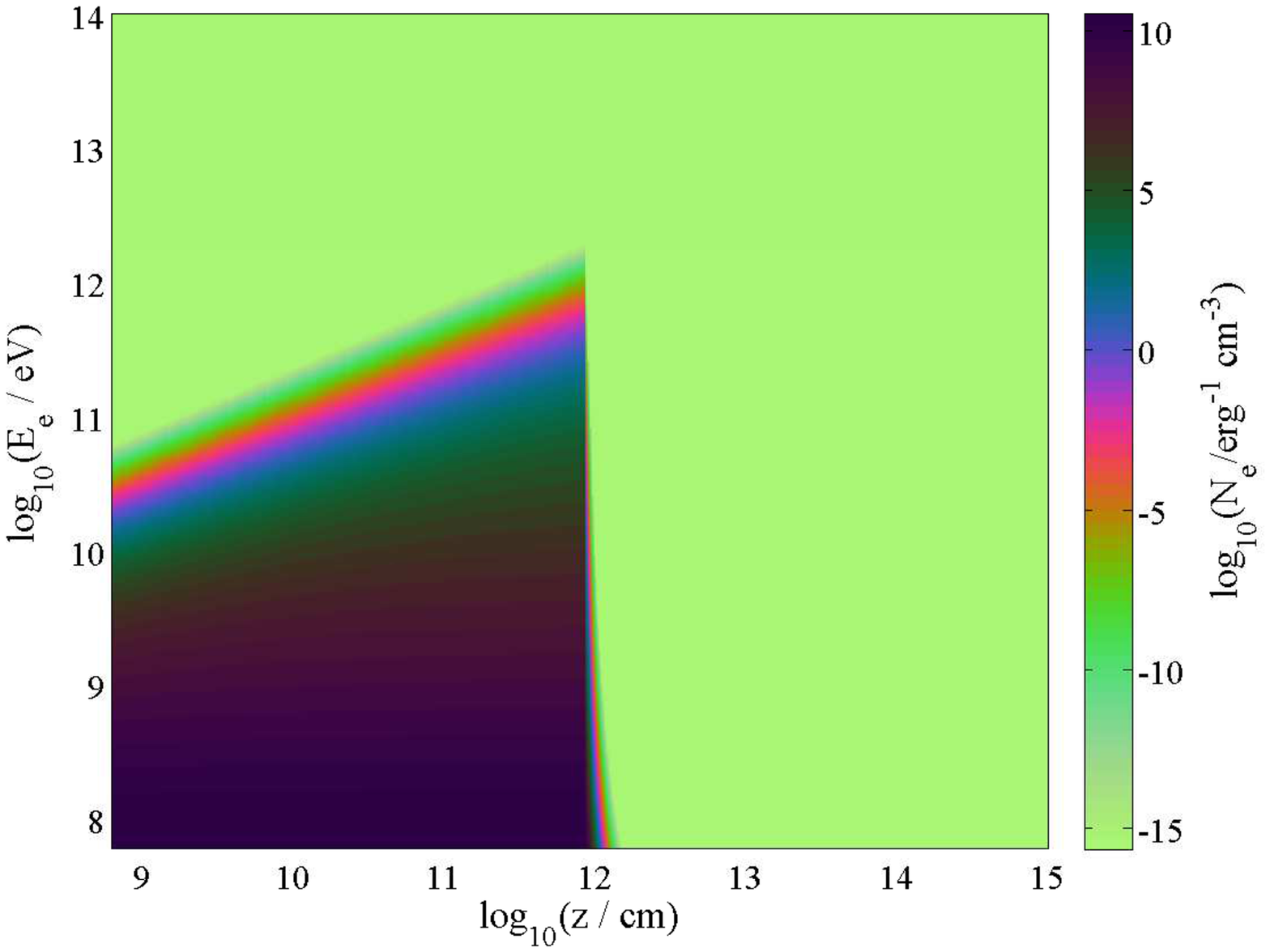}
  \includegraphics[width = 0.48\textwidth, trim=0 0 0 0, clip]{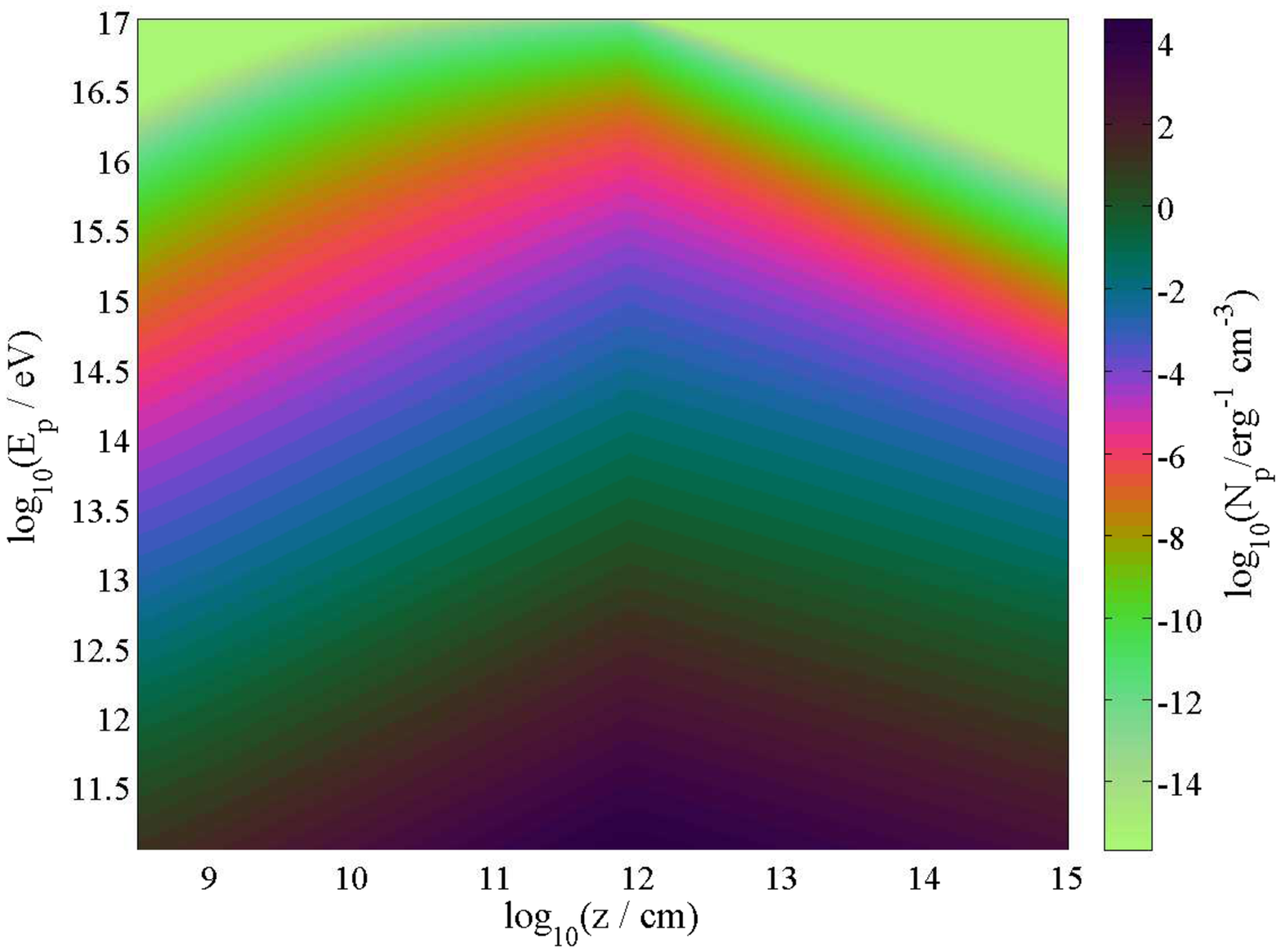}
 \caption{Steady-state distribution of primary electrons (left) and protons (right). Top panels correspond to Model A and  bottom panels to Model B. }
 \label{fig:distribution}
\end{figure*}

Figure \ref{fig:distribution} shows the particle distributions calculated from Eq. \ref{eq:transport}.  The effect of the convective term in eq. (\ref{eq:transport}) is patent from the plots: there exists a transport of 
particles from the acceleration region to the outer regions of the jet. Depending on the model, these particles can remain quite energetic and radiate far from where they are injected.

The magnetic field plays a fundamental role in the electron distribution outside the acceleration region, where the particles are only subject to cooling. In Model B the magnetic field decays slowly and electrons cool completely immediately after leaving the acceleration region, whereas in Model A a lower magnetic field implies that there is a significant number of non-thermal electrons for $z>z_{\rm max}$.

\subsection{Spectral energy distributions and absorption}
\label{sect:sed}

The best fit SEDs were introduced in Fig. \ref{fig:SEDs}; notice that the data are not simultaneous. As expected, radio emission is fitted by the synchrotron radiation of primary 
electrons in both models. Synchrotron emission of protons and secondary particles does not contribute significantly in any case. IR observations are well explained by the stellar radiation. The emission from the corona (see eq. \ref{eq:corona_spectrum}) accounts for the X-ray observations. The X-rays below $\sim 2$~keV are strongly affected by absorption interstellar medium and those data were not included in the fit.

The soft gamma-ray tail in the MeV range detected by \emph{COMPTEL} has previously been explained by \cite{romero2014} as emission from non-thermal electrons in the corona. This scenario fits into the results of Model A, where radiation in this band cannot be attributed to the jet. In Model B, however, the MeV data is well fitted by the cutoff of the synchrotron spectrum of primary electrons in the jet. In this model the synchrotron cutoff energy is higher than in Model A since the electrons radiate all their energy budget inside the acceleration region where the magnetic field is higher.

In both models the gamma-ray emission up to $\sim 100$ GeV is of leptonic origin. In Model A it is dominated by the IC radiation from the 
scattering of stellar photons, and in Model B by SSC. Finally, all the emission above $\sim 1$~TeV is completely dominated by hadronic processes. Whereas in Model A there are practically no photons above $\sim 100$ GeV, in Model B the very high energy part of the SED extends up to $\sim 100$ TeV with an approximately flat spectrum. The predicted emission is just below the upper limits of MAGIC. It is interesting to point out that in our model the levels of very high-energy gamma-ray emission (mainly $pp$) increase with the extension of the acceleration region, at fixed hadronic content. The observed upper limits thus regulate both the values of the parameters $a$ and $z_{\rm acc}$. Notice that we obtain hadronic emission up to much higher energies in Model B, even though this is the model with a smaller content of relativistic protons in the jets ($a=0.07$ vs. $a=39$ in Model A). In this model, however, protons attain larger energies and populate the whole jet; see Fig. \ref{fig:distribution}.

\begin{figure*}[htbp]
 \centering
 \includegraphics[width = 0.48\textwidth, trim=0 0 0 0, clip]{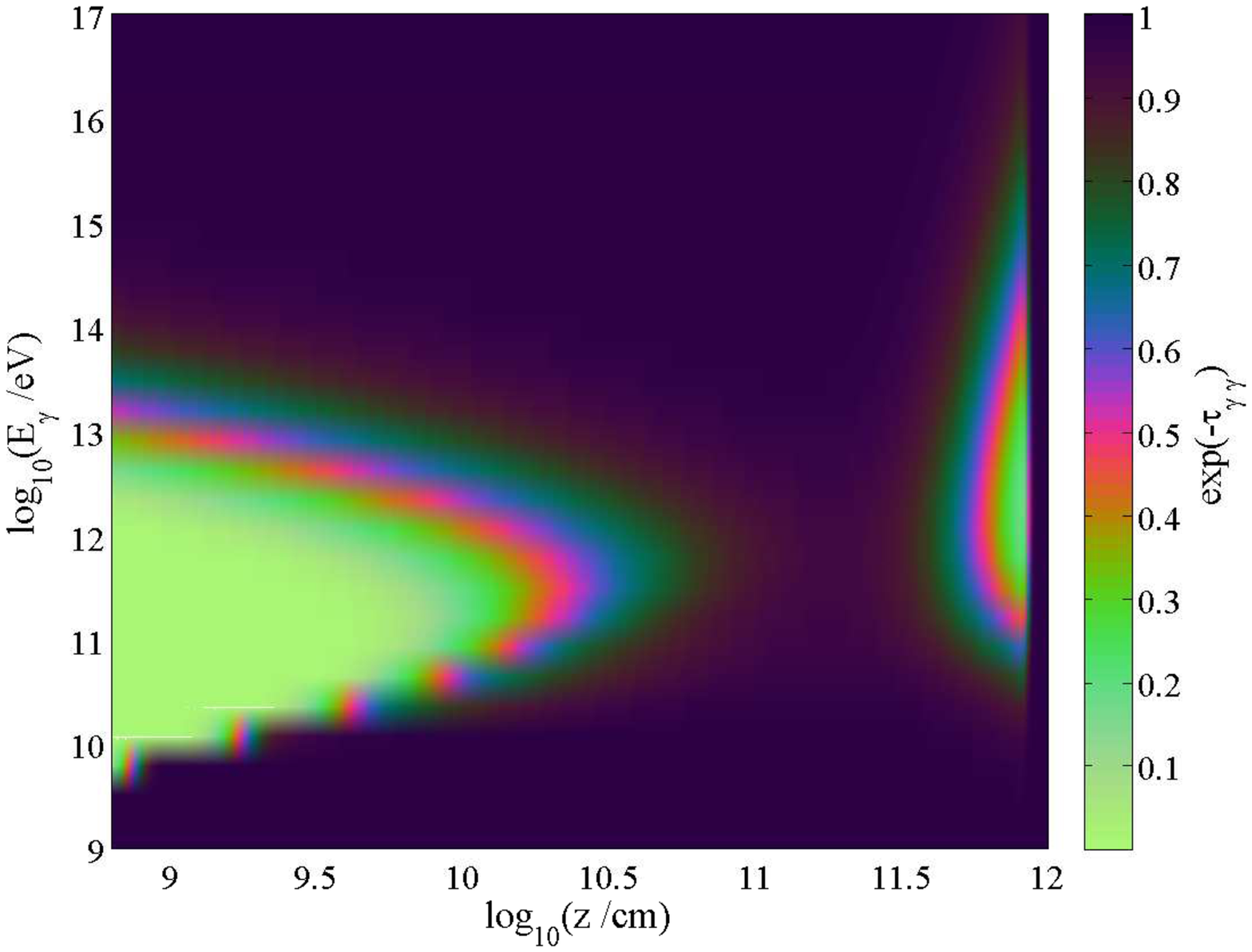}
 \includegraphics[width = 0.48\textwidth, trim=0 0 0 0, clip]{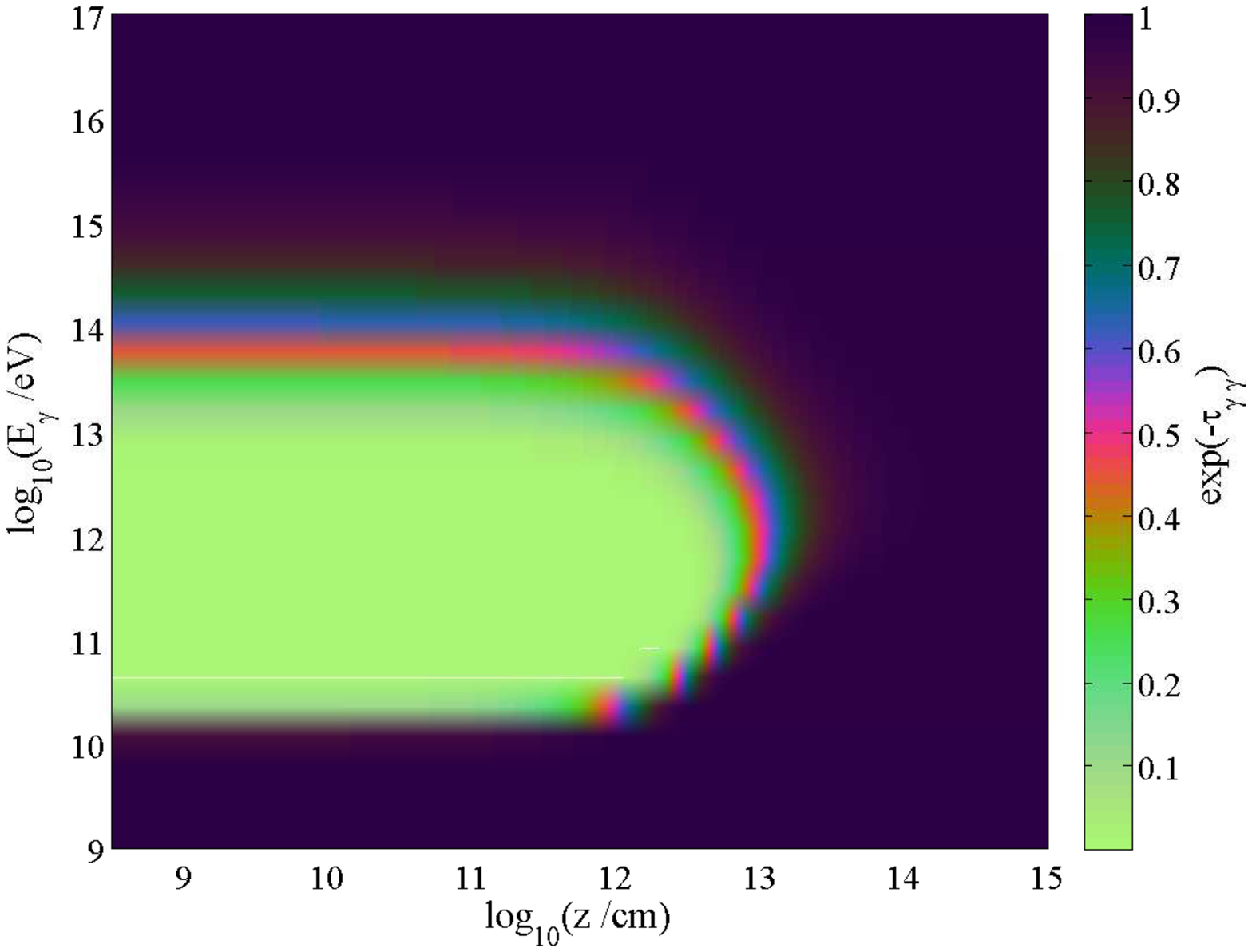}
 \caption{Attenuation factor for a photon with energy $E_\gamma$ produced at position $z$ in the jet, due to annihilation with the intrajet and external radiation fields (the same in both models). The left panel shows the contributions to the opacity of the disk radiation field at low $z$ and, for Model B, that of the intrajet synchrotron field at high $z$. The latter is negligible in Model A. The right panel shows the same for the stellar radiation field. }
 \label{fig:opacity}
\end{figure*}

The SEDs in Fig. \ref{fig:SEDs} have already been corrected by absorption applying the attenuation coefficients shown in Fig. \ref{fig:opacity}. There are three main target radiation fields: stellar photons, soft X-ray photons from the disk, and the intrajet radiation field. Since the properties of the star and the disk are the same in both models, so are the corresponding attenuation factors. In Model A internal opacity is negligible while in Model B it only adds a $\it{bump}$ at high $z$. Annihilation in the stellar photon field is the main absorption channel in both models. The high-energy radiation produced at heights on the jet larger than the binary separation is therefore unabsorbed, since the stellar radiation density is already very diluted at such distances from the star.

\subsection{Synthetic radio maps}
\label{sec:radio_maps}

From the best-fit models we produced synthetic radio maps at \mbox{8.4 GHz} for the jet of Cygnus X-1 with the aim of comparing them with the resolved radio emission detected by \cite{stirling2001}. 

We calculated the radio flux projected in the plane of the sky integrating the volumetric emissivity of the jet along the line of sight. This map was then convolved with a bidimensional Gaussian function of full width at half maximum (FWHM) of $2.25\times 0.86$ mas$^2$ to mimic the effect of an array with a beam as in Fig. 3 of \cite{stirling2001}; the chosen separation between ``pointings'' was of one beam radius in each direction. 

\begin{figure*}
\centering
 \includegraphics[width = 0.48 \textwidth, keepaspectratio]{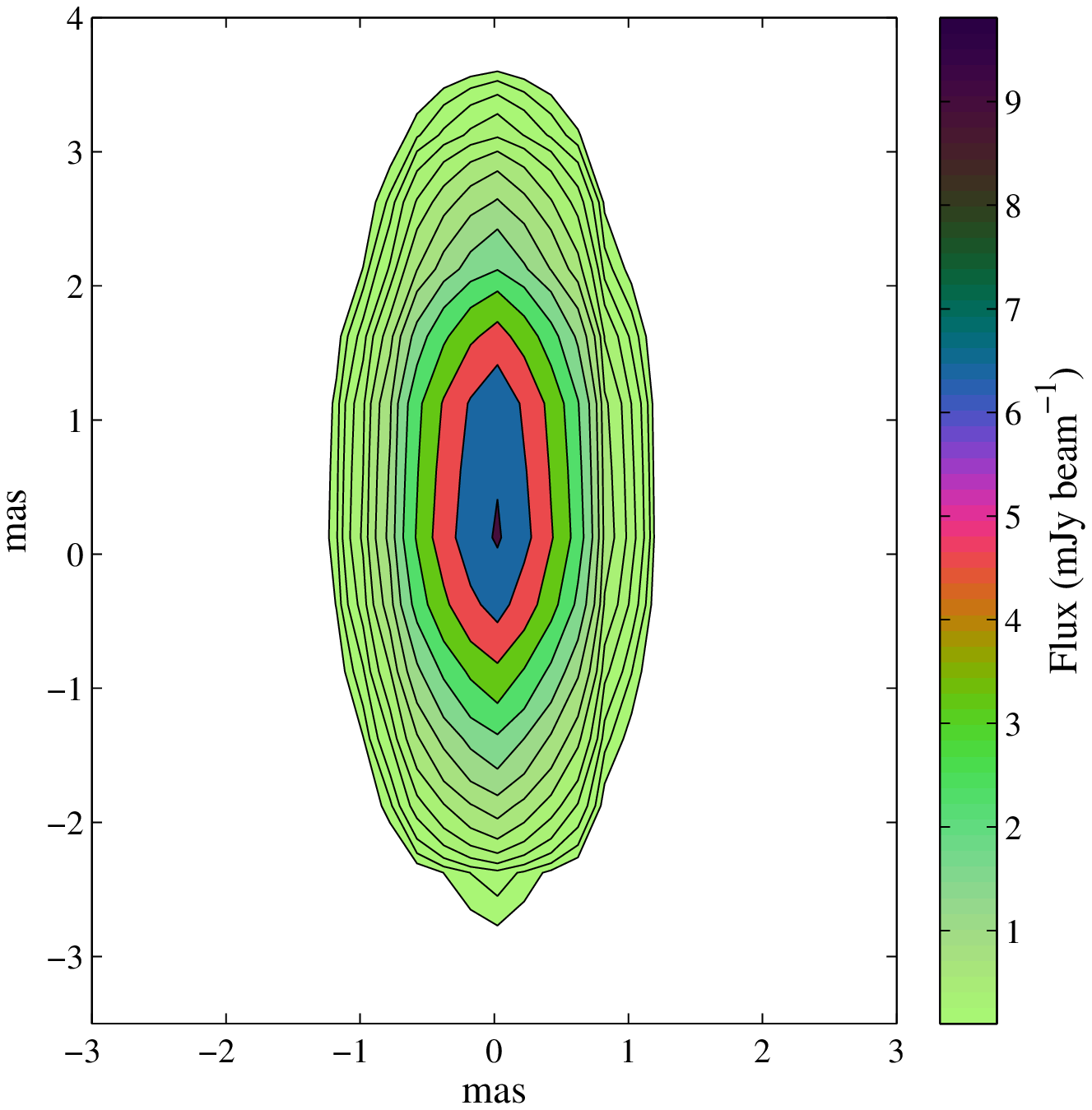}
 \includegraphics[width = 0.48 \textwidth, keepaspectratio]{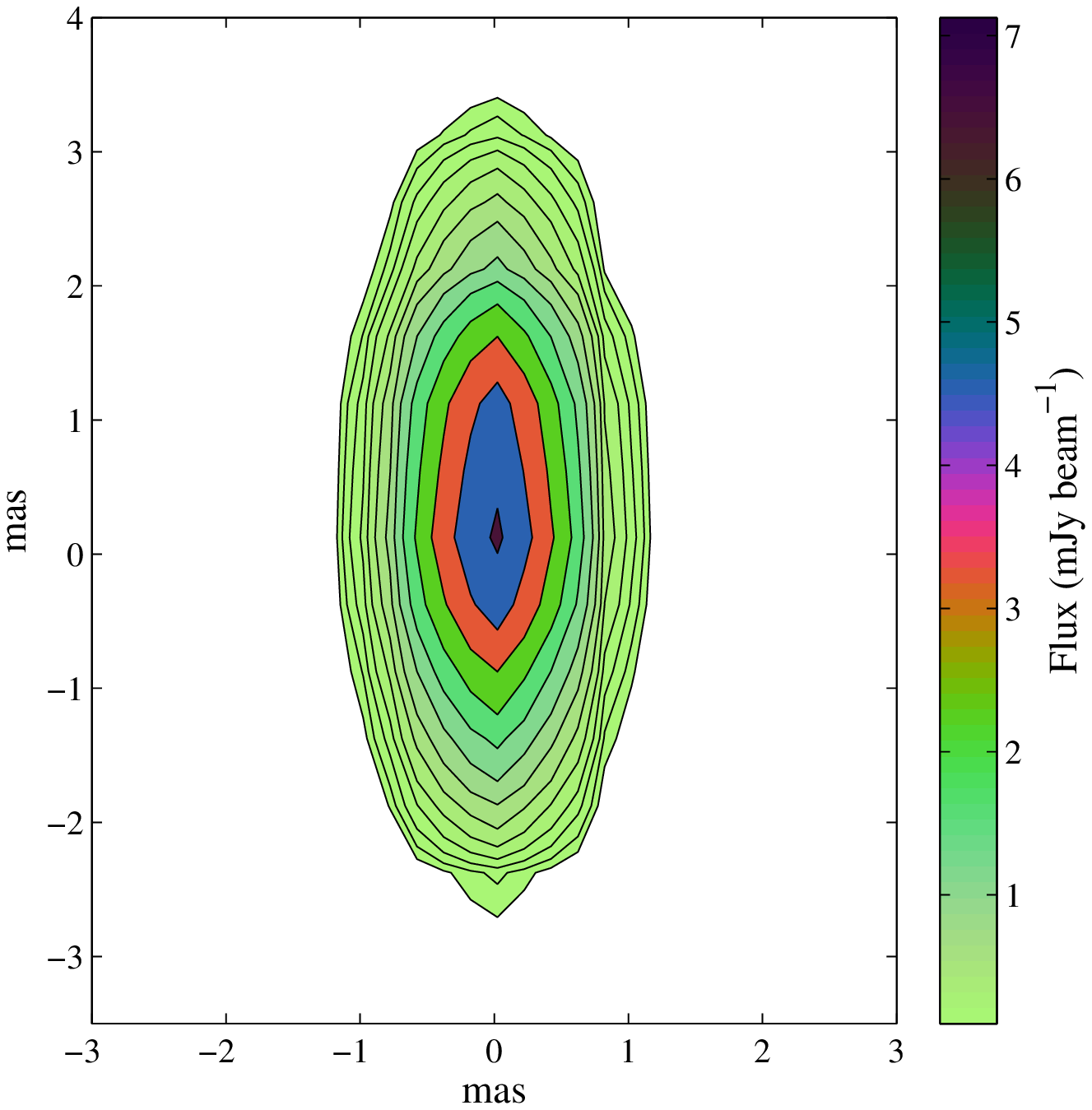}
 \caption{Image of the jet at $8.4$ GHz after convolution with a Gaussian beam of $2.25\times 0.86$ mas$^2$ for Model A (left) and Model B (right). The origin of coordinates was chosen to coincide with the position of the flux maximum. Contours are spaced in factors of $\sqrt{2}$; the lowest contour corresponds to 0.1~mJy~beam$^{-1}$.}
 \label{fig:mapas}
\end{figure*}

The results are shown in Fig. \ref{fig:mapas} for both best-fit models. The contours levels are detailed in the caption. The origin of coordinates was chosen to coincide with the maximum of the flux. The region where radio emission originates is much more extended in Model A than in Model B. This is in agreement with the electron distributions presented above: in Model A there are relativistic particles close to the end of the jet whereas in Model B they cool immediately after leaving the acceleration region ($z_{\rm{max}} \sim 8 \times 10^{11}$~cm). Nevertheless, in both models the radio emission at this frequency is confined to a small region compared to the area of the beam. Thus, the smoothed maps result very similar. 

The levels of radio emission obtained (maximum flux \mbox{$\sim 9.8$ mJy beam$^{-1}$} and \mbox{$\sim 7.1$ mJy beam$^{-1}$} in Model A and Model B, respectively) are comparable to those measured with the Very Long Baseline Array and the Very Large Array \citep[VLBA and VLA, respectively, ][]{stirling2001}, Giant Meter-wave Radio Telescope \citep[GMRT, ][]{pandey2006} and the Multi-Element Radio-Linked Interferometer Newtwork \citep[MERLIN, ][]{fender2006}. The spatial extension of the emission region we obtain is smaller than the size of $\sim 15$ mas of the extended radio source mapped by \citet{stirling2001}.  This is discussed in the next section. 

\section{Discussion}
\label{sec:discussion}

Several radiative jet models have been applied to \mbox{Cygnus X-1}. In this work, we present for the first time an analysis of the broadband SED of this source applying a lepto-hadronic, inhomogeneous jet model that features the treatment of the spatial and energy distribution of primary as well as secondary relativistic particles in an extended region. We fitted the broadband SED of Cygnus X-1 including data from the radio wavelengths to upper limits at very high energy gamma rays. We found two sets of best-fit parameters that lead to SEDs with quite different characteristics. Below we discuss the most interesting results.

The origin of the MeV tail in Cygnus X-1 is not yet clear. A possibility is that the soft gamma rays are produced in the corona. \citet{poutanen2009}, for example, were able to explain these data with a hybrid Comptonization model. Recently, \cite{romero2014} showed that if the corona contains non-thermal protons, the MeV tail may be synchrotron radiation of secondary pairs. An alternative scenario is that the MeV photons are produced in the jets. Indeed,  \cite{zdziarski2014} could explain these data as electron synchrotron emission from the jet; the GeV gamma rays, however, cannot be fit simultaneously in the same model. In our Model B the MeV tail is also fitted with electron synchrotron radiation, but we are able to reproduce the \emph{Fermi}-LAT data and upper limits as well. The detection of polarized hard X-ray/soft $\gamma$-ray emission with \textit{INTEGRAL} \citep{laurent2011, jourdain2012} suggests that it might originate in the jets, since it is not observed during the high/soft state \citep{
rodriguez2015}. New insights into this question may be gained by modeling the polarization of the synchrotron and IC emission in the jet. This requires a more detailed description of the large-scale magnetic field of the jet, and is left for a forthcoming work. 
 
In our model the GeV gamma rays are of leptonic origin, namely IC scattering of stellar photons (Model A) or synchrotron self-Compton (Model B). While respecting the upper limits of MAGIC, the calculated TeV emission in Model B is of the order of $\sim 10^{32}$ erg s$^{-1}$; it might be marginally detectable in the future with CTA depending on the final sensitivity of the array. An interesting feature of our model is that the level of very high-energy hadronic emission grows as the acceleration region is extended. The observational upper limits at $\gtrsim 1$ GeV, then, constrain not only the hadronic content of the jet but also the size of the acceleration site.

We have also calculated the radio image of the jet attempting to reproduce that obtained by \citet{stirling2001} with the VLBA at $8.4$ GHz.  The flux levels predicted in our model are in good agreement with the observations. After smoothing with a Gaussian beam of $2.25\times 0.86$ mas$^2$ the size of the synthetic source is similar in both models, but smaller than the observed one (notice, however, these observations of are not simultaneous with any of those included in the fits of the SED). The reason is that in both best-fit models the bulk of the electron synchrotron radiation is emitted in a relatively thin region of the jet compared with its total length. 
Furthermore, given that the beam size is comparable to that of the emission region, both models lead to similar extensions of the convolved radio maps. The discrepancy we find in the extent of the radio emission region points to a deficiency in our physical assumptions, and thus our modeling, of at least two issues: the size and/or location of the acceleration zone(s) and the morphology of the magnetic field. Sites of particle acceleration (and thus radio emission) may exist farther away from the black hole than we assumed here if shocks develop in the outflow. This is indeed very likely to occur. It is well known, for instance, that jets are subject to recollimation shocks; these have been observed in extragalactic outflows and appear in numerical simulations \citep[e.g.][]{perucho2010}. In Cygnus X-1 in particular a bend or kink in the jet at $\sim 7$ mas from the core was reported by \citet{stirling2001}; it disappears on a time-scale of $\leq 2$~d. This bend may be related to the structure of the 
magnetic field but also to the impact of the stellar wind on the jet \citep[see e.g.][]{perucho2012, YoonHeinz2015}. In any case this bend could be a region with suitable conditions for particle re-acceleration. Shocked regions may also developed in the jets if they are traversed by clumps in the wind \citep{araudo2009}. 

Throughout this work we have assumed that the binary is at superior conjunction.  This is the configuration for which the absorption of gamma rays in the stellar photon field is expected to be maximum. 
In our best-fit models, however, the radiation of secondary pairs injected by two-photon annihilation does not contribute significantly to the radiative output of the source (i.e., the synchrotron luminosity of these pairs lies below $10^{28}$~erg~s$^{-1}$). Furthermore, the effect of absorption on the total SED is almost negligible since the high-energy radiation is produced above the region where absorption is significant. Thus we do not study the dependence of our results on the orbital configuration. We remark that our results for the absorption coefficient at superior conjunction are in agreement with those of \cite{romero2010}. 

\section{Concluding remarks}
\label{sec:conclusion}

In this article we applied an inhomogeneous, lepto-hadronic radiative jet model to study the broadband emission of \mbox{Cygnus X-1}. We obtain two SEDs that fit the available observational data and upper limits from radio wavelengths to TeV energies. The main difference between the two best-fit models lies in the predictions for the gamma-ray band, where the origin of the emission is not yet settled. 

In particular, in Model B, where the particle injection is harder and the magnetic field decays more slowly, we are able to fit the MeV tail with electron synchrotron emission. A similar result was previously obtained by \cite{zdziarski2014}, but our best SED simultaneously fits the MeV and the GeV data. When a softer particle injection and a rapidly decaying magnetic field is considered (Model A), the same data cannot be explained as radiation from the jet in our model. In this scenario non-thermal radiation from a corona remains as an alternative to account for the soft gamma-ray emission \citep{poutanen2009, romero2014}. Further measurements together with a detailed modeling of the magnetic field and the X-ray polarization should be useful to reveal the origin of the MeV tail in Cygnus X-1. 

The very high energy emission in our model is of hadronic origin. The most relevant interaction are $pp$ collisions with the matter in the stellar wind as target; the gamma rays are hardly affected by absorption. In Model B the emission above $\sim 1~$TeV is very close to the upper limits of MAGIC, and  might be detectable with the next generation of gamma-ray telescopes such as CTA.

As usual, in our model the radio emission is well explained as electron synchrotron radiation from the jet. In this work we have gone one step further by attempting to reproduce the morphology of the radio jet at 8.4 GHz as mapped by \cite{stirling2001}. For both best-fit SEDs we calculated the radio image of the jet on the plane of the sky, and simulated instrumental effects by convolving them with a $2.25\times 0.86$ mas$^2$ Gaussian beam. Although we obtain flux levels in agreement with those reported, the synthetic radio source is more compact. The fundamental cause is that the our best-fit models of the SED favour a compact acceleration region where most of the electron synchrotron emission is concentrated. The parameter that largely determines the properties of the electron distribution along the jet is the magnetic field, so a detailed modeling of this aspect appears  once again necessary.  Non-thermal emission on large spatial scales may also occur if there are several sites of particle acceleration 
along the jets. Acceleration regions are usually associated to shock fronts, and these are known to develop frequently in outflows.  

The quality and variety of the available observational data thus makes it timely to progress towards the introduction  of some aspects of the large-scale jet dynamics in radiative models.  Reproducing the morphology of the source and the polarization of the non-thermal radiation may prove useful tools in this direction. Furthermore, such information may help to break some of the degeneracy in the models that remains after fitting the SED. We will address these issues in forthcoming works.

\begin{acknowledgements}
We thank Paula Benaglia, Valent\'i Bosch-Ramon, Nicol\'as Casco, Manuel Fern\'andez L\'opez and Cintia Peri for their help with the radio maps. This work was supported by grants PICT 2012-00878 from ANPCyT (Argentina) and AYA 2013-47447-C3-1-P from MEyC (Spain). 
\end{acknowledgements}

\bibliographystyle{aa}

\def\apj{ApJ}
\def\apjs{ApJS}
\def\apjl{ApJ}
\def\aj{AJ}
\def\mnras{MNRAS}
\def\aa{A\&A}
\def\nat{Nature}
\def\araa{ARA\&A}
\def\aap{A\&A}

\bibliography{article_Pepe}

\begin{thebibliography}{69}
\expandafter\ifx\csname natexlab\endcsname\relax\def\natexlab#1{#1}\fi

\bibitem[{{Aharonian}(2004)}]{aharonian2004}
{Aharonian}, F.~A. 2004, {Very high energy cosmic gamma radiation: a crucial
  window on the extreme Universe}

\bibitem[{{Albert} {et~al.}(2007){Albert}, {Aliu}, {Anderhub}, {Antoranz},
  {Armada}, {Baixeras}, {Barrio}, {Bartko}, {Bastieri}, {Becker}, {Bednarek},
  {Berger}, {Bigongiari}, {Biland}, {Bock}, {Bordas}, {Bosch-Ramon}, {Bretz},
  {Britvitch}, {Camara}, {Carmona}, {Chilingarian}, {Coarasa}, {Commichau},
  {Contreras}, {Cortina}, {Costado}, {Curtef}, {Danielyan}, {Dazzi}, {De
  Angelis}, {Delgado}, {de los Reyes}, {De Lotto}, {Domingo-Santamar{\'{\i}}a},
  {Dorner}, {Doro}, {Errando}, {Fagiolini}, {Ferenc}, {Fern{\'a}ndez}, {Firpo},
  {Flix}, {Fonseca}, {Font}, {Fuchs}, {Galante}, {Garc{\'{\i}}a-L{\'o}pez},
  {Garczarczyk}, {Gaug}, {Giller}, {Goebel}, {Hakobyan}, {Hayashida},
  {Hengstebeck}, {Herrero}, {H{\"o}hne}, {Hose}, {Hsu}, {Jacon}, {Jogler},
  {Kosyra}, {Kranich}, {Kritzer}, {Laille}, {Lindfors}, {Lombardi}, {Longo},
  {L{\'o}pez}, {L{\'o}pez}, {Lorenz}, {Majumdar}, {Maneva}, {Mannheim},
  {Mansutti}, {Mariotti}, {Mart{\'{\i}}nez}, {Mazin}, {Merck}, {Meucci},
  {Meyer}, {Miranda}, {Mirzoyan}, {Mizobuchi}, {Moralejo}, {Nieto}, {Nilsson},
  {Ninkovic}, {O{\~n}a-Wilhelmi}, {Otte}, {Oya}, {Panniello}, {Paoletti},
  {Paredes}, {Pasanen}, {Pascoli}, {Pauss}, {Pegna}, {Persic}, {Peruzzo},
  {Piccioli}, {Prandini}, {Puchades}, {Raymers}, {Rhode}, {Rib{\'o}}, {Rico},
  {Rissi}, {Robert}, {R{\"u}gamer}, {Saggion}, {Saito}, {S{\'a}nchez},
  {Sartori}, {Scalzotto}, {Scapin}, {Schmitt}, {Schweizer}, {Shayduk},
  {Shinozaki}, {Shore}, {Sidro}, {Sillanp{\"a}{\"a}}, {Sobczynska}, {Stamerra},
  {Stark}, {Takalo}, {Temnikov}, {Tescaro}, {Teshima}, {Torres}, {Turini},
  {Vankov}, {Vitale}, {Wagner}, {Wibig}, {Wittek}, {Zandanel}, {Zanin}, \&
  {Zapatero}}]{albert2007}
{Albert}, J., {Aliu}, E., {Anderhub}, H., {et~al.} 2007, \apjl, 665, L51

\bibitem[{{Araudo} {et~al.}(2009){Araudo}, {Bosch-Ramon}, \&
  {Romero}}]{araudo2009}
{Araudo}, A.~T., {Bosch-Ramon}, V., \& {Romero}, G.~E. 2009, \aap, 503, 673

\bibitem[{{Becker} \& {Kafatos}(1995)}]{becker1995}
{Becker}, P.~A. \& {Kafatos}, M. 1995, \apj, 453, 83

\bibitem[{{Bednarek}(2005)}]{bednarek2005}
{Bednarek}, W. 2005, \apj, 631, 466

\bibitem[{{Belloni} {et~al.}(1996){Belloni}, {Mendez}, {van der Klis},
  {Hasinger}, {Lewin}, \& {van Paradijs}}]{belloni1996}
{Belloni}, T., {Mendez}, M., {van der Klis}, M., {et~al.} 1996, \apjl, 472,
  L107

\bibitem[{{Bodaghee} {et~al.}(2013){Bodaghee}, {Tomsick}, {Pottschmidt},
  {Rodriguez}, {Wilms}, \& {Pooley}}]{bodaghee-etal2013}
{Bodaghee}, A., {Tomsick}, J.~A., {Pottschmidt}, K., {et~al.} 2013, \apj, 775,
  98

\bibitem[{{Bosch-Ramon} {et~al.}(2006){Bosch-Ramon}, {Romero}, \&
  {Paredes}}]{boschRamon2006}
{Bosch-Ramon}, V., {Romero}, G.~E., \& {Paredes}, J.~M. 2006, \aap, 447, 263

\bibitem[{{Brocksopp} {et~al.}(2002){Brocksopp}, {Fender}, \&
  {Pooley}}]{brocksopp2002}
{Brocksopp}, C., {Fender}, R.~P., \& {Pooley}, G.~G. 2002, \mnras, 336, 699

\bibitem[{{Bulgarelli} {et~al.}(2010){Bulgarelli}, {Pittori}, {Lucarelli},
  {Striani}, {Gianotti}, {Trifoglio}, {Sabatini}, {Tavani}, {Verrecchia},
  {Trois}, {Chen}, {Giuliani}, {Mereghetti}, {Caraveo}, {Perotti},
  {Vercellone}, {D'Ammando}, {Donnarumma}, {Del Monte}, {Evangelista},
  {Feroci}, {Lazzarotto}, {Pacciani}, {Soffitta}, {Costa}, {Lapshov},
  {Rapisarda}, {Argan}, {Piano}, {Pucella}, {Vittorini}, {Fuschino}, {Galli},
  {Labanti}, {Marisaldi}, {Di Cocco}, {Pellizzoni}, {Pilia}, {Barbiellini},
  {Longo}, {Moretti}, {Vallazza}, {Morselli}, {Picozza}, {Prest}, {Lipari},
  {Zanello}, {Cattaneo}, {Rappoldi}, {Santolamazza}, {Colafrancesco}, {Giommi},
  \& {Salotti}}]{bulgarelli-etal2010}
{Bulgarelli}, A., {Pittori}, C., {Lucarelli}, F., {et~al.} 2010, The
  Astronomer's Telegram, 2512, 1

\bibitem[{{Cadolle Bel} {et~al.}(2006){Cadolle Bel}, {Sizun}, {Goldwurm},
  {Rodriguez}, {Laurent}, {Zdziarski}, {Foschini}, {Goldoni}, {Gouiff{\`e}s},
  {Malzac}, {Jourdain}, \& {Roques}}]{cadolleBell-etal2006}
{Cadolle Bel}, M., {Sizun}, P., {Goldwurm}, A., {et~al.} 2006, \aap, 446, 591

\bibitem[{{Dermer} \& {Schlickeiser}(1993)}]{dermer1993}
{Dermer}, C.~D. \& {Schlickeiser}, R. 1993, \apj, 416, 458

\bibitem[{{Di Salvo} {et~al.}(2001){Di Salvo}, {Done}, {{\.Z}ycki}, {Burderi},
  \& {Robba}}]{diSalvo2001}
{Di Salvo}, T., {Done}, C., {{\.Z}ycki}, P.~T., {Burderi}, L., \& {Robba},
  N.~R. 2001, \apj, 547, 1024

\bibitem[{{D{\'{\i}}az Trigo} {et~al.}(2013){D{\'{\i}}az Trigo},
  {Miller-Jones}, {Migliari}, {Broderick}, \& {Tzioumis}}]{diazTrigo2013}
{D{\'{\i}}az Trigo}, M., {Miller-Jones}, J.~C.~A., {Migliari}, S., {Broderick},
  J.~W., \& {Tzioumis}, T. 2013, \nat, 504, 260

\bibitem[{{Dove} {et~al.}(1997){Dove}, {Wilms}, {Maisack}, \&
  {Begelman}}]{dove-etal1997}
{Dove}, J.~B., {Wilms}, J., {Maisack}, M., \& {Begelman}, M.~C. 1997, \apj,
  487, 759

\bibitem[{{Fender} {et~al.}(2000){Fender}, {Pooley}, {Durouchoux}, {Tilanus},
  \& {Brocksopp}}]{fender2000}
{Fender}, R.~P., {Pooley}, G.~G., {Durouchoux}, P., {Tilanus}, R.~P.~J., \&
  {Brocksopp}, C. 2000, \mnras, 312, 853

\bibitem[{{Fender} {et~al.}(2006){Fender}, {Stirling}, {Spencer}, {Brown},
  {Pooley}, {Muxlow}, \& {Miller-Jones}}]{fender2006}
{Fender}, R.~P., {Stirling}, A.~M., {Spencer}, R.~E., {et~al.} 2006, \mnras,
  369, 603

\bibitem[{{Gallo} {et~al.}(2005){Gallo}, {Fender}, {Kaiser}, {Russell},
  {Morganti}, {Oosterloo}, \& {Heinz}}]{gallo2005}
{Gallo}, E., {Fender}, R., {Kaiser}, C., {et~al.} 2005, \nat, 436, 819

\bibitem[{{Ghisellini} {et~al.}(1985){Ghisellini}, {Maraschi}, \&
  {Treves}}]{Ghisellini85}
{Ghisellini}, G., {Maraschi}, L., \& {Treves}, A. 1985, 146, 204

\bibitem[{{Heinz}(2006)}]{Heinz2006}
{Heinz}, S. 2006, \apj, 636, 316

\bibitem[{{Jourdain} {et~al.}(2012){Jourdain}, {Roques}, {Chauvin}, \&
  {Clark}}]{jourdain2012}
{Jourdain}, E., {Roques}, J.~P., {Chauvin}, M., \& {Clark}, D.~J. 2012, \apj,
  761, 27

\bibitem[{{Khangulyan} {et~al.}(2008){Khangulyan}, {Aharonian}, \&
  {Bosch-Ramon}}]{khangulyan2008}
{Khangulyan}, D., {Aharonian}, F., \& {Bosch-Ramon}, V. 2008, \mnras, 383, 467

\bibitem[{{Khangulyan} {et~al.}(2014){Khangulyan}, {Aharonian}, \&
  {Kelner}}]{khangulyan2014}
{Khangulyan}, D., {Aharonian}, F.~A., \& {Kelner}, S.~R. 2014, \apj, 783, 100

\bibitem[{{Lachowicz} {et~al.}(2006){Lachowicz}, {Zdziarski},
  {Schwarzenberg-Czerny}, {Pooley}, \& {Kitamoto}}]{lachowicz-etal2006}
{Lachowicz}, P., {Zdziarski}, A.~A., {Schwarzenberg-Czerny}, A., {Pooley},
  G.~G., \& {Kitamoto}, S. 2006, \mnras, 368, 1025

\bibitem[{{Lamers} \& {Cassinelli}(1990)}]{lamers1990}
{Lamers}, H.~J.~G.~L.~M. \& {Cassinelli}, J.~P. 1990, {Introduction to stellar
  winds}

\bibitem[{{Laurent} {et~al.}(2011){Laurent}, {Rodriguez}, {Wilms}, {Cadolle
  Bel}, {Pottschmidt}, \& {Grinberg}}]{laurent2011}
{Laurent}, P., {Rodriguez}, J., {Wilms}, J., {et~al.} 2011, Science, 332, 438

\bibitem[{{Levinson} \& {Waxman}(2001)}]{LevinsonWaxman2001}
{Levinson}, A. \& {Waxman}, E. 2001, Physical Review Letters, 87, 171101

\bibitem[{{Lind} \& {Blandford}(1985)}]{Lind1985}
{Lind}, K.~R. \& {Blandford}, R.~D. 1985, \apj, 295, 358

\bibitem[{{Malyshev} {et~al.}(2013){Malyshev}, {Zdziarski}, \&
  {Chernyakova}}]{malyshev2013}
{Malyshev}, D., {Zdziarski}, A.~A., \& {Chernyakova}, M. 2013, \mnras, 434,
  2380

\bibitem[{{McConnell} {et~al.}(2002{\natexlab{a}}){McConnell}, {Zdziarski},
  {Bennett}, {Bloemen}, {Collmar}, {Hermsen}, {Kuiper}, {Paciesas}, {Phlips},
  {Poutanen}, {Ryan}, {Sch{\"o}nfelder}, {Steinle}, \&
  {Strong}}]{mcConnell2000}
{McConnell}, M.~L., {Zdziarski}, A.~A., {Bennett}, K., {et~al.}
  2002{\natexlab{a}}, \apj, 572, 984

\bibitem[{{McConnell} {et~al.}(2002{\natexlab{b}}){McConnell}, {Zdziarski},
  {Bennett}, {Bloemen}, {Collmar}, {Hermsen}, {Kuiper}, {Paciesas}, {Phlips},
  {Poutanen}, {Ryan}, {Sch{\"o}nfelder}, {Steinle}, \&
  {Strong}}]{mcConnell2002}
{McConnell}, M.~L., {Zdziarski}, A.~A., {Bennett}, K., {et~al.}
  2002{\natexlab{b}}, \apj, 572, 984

\bibitem[{{Migliari} {et~al.}(2002){Migliari}, {Fender}, \&
  {M{\'e}ndez}}]{migliari2002}
{Migliari}, S., {Fender}, R., \& {M{\'e}ndez}, M. 2002, Science, 297, 1673

\bibitem[{{Mirabel} {et~al.}(1996){Mirabel}, {Claret}, {Cesarsky}, {Boulade},
  \& {Cesarsky}}]{mirabel1996}
{Mirabel}, I.~F., {Claret}, A., {Cesarsky}, C.~J., {Boulade}, O., \&
  {Cesarsky}, D.~A. 1996, \aap, 315, L113

\bibitem[{{Mirabel} {et~al.}(1998){Mirabel}, {Dhawan}, {Chaty}, {Rodriguez},
  {Marti}, {Robinson}, {Swank}, \& {Geballe}}]{mirabel1998}
{Mirabel}, I.~F., {Dhawan}, V., {Chaty}, S., {et~al.} 1998, \aap, 330, L9

\bibitem[{{Orosz} {et~al.}(2011){Orosz}, {McClintock}, {Aufdenberg},
  {Remillard}, {Reid}, {Narayan}, \& {Gou}}]{orosz2011}
{Orosz}, J.~A., {McClintock}, J.~E., {Aufdenberg}, J.~P., {et~al.} 2011, \apj,
  742, 84

\bibitem[{{Pandey} {et~al.}(2006){Pandey}, {Rao}, {Pooley}, {Durouchoux},
  {Manchanda}, \& {Ishwara-Chandra}}]{pandey2006}
{Pandey}, M., {Rao}, A.~P., {Pooley}, G.~G., {et~al.} 2006, \aap, 447, 525

\bibitem[{{Persi} {et~al.}(1980){Persi}, {Ferrari-Toniolo}, {Grasdalen}, \&
  {Spada}}]{persi1980}
{Persi}, P., {Ferrari-Toniolo}, M., {Grasdalen}, G.~L., \& {Spada}, G. 1980,
  \aap, 92, 238

\bibitem[{{Perucho} \& {Bosch-Ramon}(2012)}]{perucho2012}
{Perucho}, M. \& {Bosch-Ramon}, V. 2012, \aap, 539, A57

\bibitem[{{Perucho} {et~al.}(2010){Perucho}, {Bosch-Ramon}, \&
  {Khangulyan}}]{perucho2010}
{Perucho}, M., {Bosch-Ramon}, V., \& {Khangulyan}, D. 2010, \aap, 512, L4

\bibitem[{{Poutanen}(1998)}]{poutanen1998}
{Poutanen}, J. 1998, in Theory of Black Hole Accretion Disks, ed. M.~A.
  {Abramowicz}, G.~{Bj{\"o}rnsson}, \& J.~E. {Pringle}, 100--122

\bibitem[{{Poutanen} {et~al.}(1997){Poutanen}, {Krolik}, \&
  {Ryde}}]{poutanen1997}
{Poutanen}, J., {Krolik}, J.~H., \& {Ryde}, F. 1997, \mnras, 292, L21

\bibitem[{{Poutanen} \& {Vurm}(2009)}]{poutanen2009}
{Poutanen}, J. \& {Vurm}, I. 2009, \apjl, 690, L97

\bibitem[{{Rahoui} {et~al.}(2011){Rahoui}, {Lee}, {Heinz}, {Hines},
  {Pottschmidt}, {Wilms}, \& {Grinberg}}]{rahoui2011}
{Rahoui}, F., {Lee}, J.~C., {Heinz}, S., {et~al.} 2011, \apj, 736, 63

\bibitem[{{Reid} {et~al.}(2011){Reid}, {McClintock}, {Narayan}, {Gou},
  {Remillard}, \& {Orosz}}]{reid2011}
{Reid}, M.~J., {McClintock}, J.~E., {Narayan}, R., {et~al.} 2011, in Bulletin
  of the American Astronomical Society, Vol.~43, American Astronomical Society
  Meeting Abstracts 217, 223.01

\bibitem[{{Reynoso} \& {Romero}(2009)}]{reynoso2009}
{Reynoso}, M.~M. \& {Romero}, G.~E. 2009, \aap, 493, 1

\bibitem[{{Rieger} {et~al.}(2007){Rieger}, {Bosch-Ramon}, \&
  {Duffy}}]{rieger2007}
{Rieger}, F.~M., {Bosch-Ramon}, V., \& {Duffy}, P. 2007, \apss, 309, 119

\bibitem[{{Rodriguez} {et~al.}(2015){Rodriguez}, {Grinberg}, {Laurent},
  {Cadolle Bel}, {Pottschmidt}, {Pooley}, {Bodaghee}, {Wilms}, \&
  {Gouiff{\`e}s}}]{rodriguez2015}
{Rodriguez}, J., {Grinberg}, V., {Laurent}, P., {et~al.} 2015, ArXiv e-prints

\bibitem[{{Romero} {et~al.}(2005){Romero}, {Christiansen}, \&
  {Orellana}}]{romero-etal2005}
{Romero}, G.~E., {Christiansen}, H.~R., \& {Orellana}, M. 2005, \apj, 632, 1093

\bibitem[{{Romero} {et~al.}(2010){Romero}, {Del Valle}, \&
  {Orellana}}]{romero2010}
{Romero}, G.~E., {Del Valle}, M.~V., \& {Orellana}, M. 2010, \aap, 518, A12

\bibitem[{{Romero} {et~al.}(2003){Romero}, {Torres}, {Kaufman Bernad{\'o}}, \&
  {Mirabel}}]{romero2003}
{Romero}, G.~E., {Torres}, D.~F., {Kaufman Bernad{\'o}}, M.~M., \& {Mirabel},
  I.~F. 2003, \aap, 410, L1

\bibitem[{{Romero} {et~al.}(2014){Romero}, {Vieyro}, \& {Chaty}}]{romero2014}
{Romero}, G.~E., {Vieyro}, F.~L., \& {Chaty}, S. 2014, \aap, 562, L7

\bibitem[{{Romero} \& {Vila}(2008)}]{romero-vila2008}
{Romero}, G.~E. \& {Vila}, G.~S. 2008, \aap, 485, 623

\bibitem[{{Rushton} {et~al.}(2011){Rushton}, {Miller-Jones}, {Paragi},
  {Maccarone}, {Pooley}, {Tudose}, {Fender}, {Spencer}, {Dhawan}, \&
  {Garrett}}]{Rushton-etal2011}
{Rushton}, A., {Miller-Jones}, J., {Paragi}, Z., {et~al.} 2011, ArXiv e-prints

\bibitem[{{Rushton} {et~al.}(2012){Rushton}, {Miller-Jones}, {Campana},
  {Evangelista}, {Paragi}, {Maccarone}, {Pooley}, {Tudose}, {Fender},
  {Spencer}, \& {Dhawan}}]{Rushton-etal2012}
{Rushton}, A., {Miller-Jones}, J.~C.~A., {Campana}, R., {et~al.} 2012, \mnras,
  419, 3194

\bibitem[{{Russell} \& {Shahbaz}(2014)}]{RussellShahbaz2014}
{Russell}, D.~M. \& {Shahbaz}, T. 2014, \mnras, 438, 2083

\bibitem[{{Rybicki} \& {Lightman}(1986)}]{RybickiLightman1986}
{Rybicki}, G.~B. \& {Lightman}, A.~P. 1986, {Radiative Processes in
  Astrophysics}

\bibitem[{{Sabatini} {et~al.}(2013){Sabatini}, {Tavani}, {Coppi}, {Pooley},
  {Del Santo}, {Campana}, {Chen}, {Evangelista}, {Piano}, {Bulgarelli},
  {Cattaneo}, {Colafrancesco}, {Del Monte}, {Giuliani}, {Giusti}, {Longo},
  {Morselli}, {Pellizzoni}, {Pilia}, {Striani}, {Trifoglio}, \&
  {Vercellone}}]{sabatini2013}
{Sabatini}, S., {Tavani}, M., {Coppi}, P., {et~al.} 2013, \apj, 766, 83

\bibitem[{{Sabatini} {et~al.}(2010){Sabatini}, {Tavani}, {Striani},
  {Bulgarelli}, {Vittorini}, {Piano}, {Del Monte}, {Feroci}, {de Pasquale},
  {Trifoglio}, {Gianotti}, {Argan}, {Barbiellini}, {Caraveo}, {Cattaneo},
  {Chen}, {D'Ammando}, {Costa}, {De Paris}, {Di Cocco}, {Donnarumma},
  {Evangelista}, {Ferrari}, {Fiorini}, {Fuschino}, {Galli}, {Giuliani},
  {Giusti}, {Labanti}, {Lazzarotto}, {Lipari}, {Longo}, {Marisaldi},
  {Mereghetti}, {Morelli}, {Moretti}, {Morselli}, {Pacciani}, {Pellizzoni},
  {Perotti}, {Picozza}, {Pilia}, {Pucella}, {Prest}, {Rapisarda}, {Rappoldi},
  {Rubini}, {Scalise}, {Soffitta}, {Trois}, {Vallazza}, {Vercellone}, {Zambra},
  {Zanello}, {Pittori}, {Verrecchia}, {Santolamazza}, {Giommi},
  {Colafrancesco}, {Antonelli}, \& {Salotti}}]{sabatini2010}
{Sabatini}, S., {Tavani}, M., {Striani}, E., {et~al.} 2010, \apjl, 712, L10

\bibitem[{{Sell} {et~al.}(2015){Sell}, {Heinz}, {Richards}, {Maccarone},
  {Russell}, {Gallo}, {Fender}, {Markoff}, \& {Nowak}}]{sell2015}
{Sell}, P.~H., {Heinz}, S., {Richards}, E., {et~al.} 2015, \mnras, 446, 3579

\bibitem[{{Shakura} \& {Sunyaev}(1973)}]{ShakuraSunyaev1973}
{Shakura}, N.~I. \& {Sunyaev}, R.~A. 1973, \aap, 24, 337

\bibitem[{{Stirling} {et~al.}(2001){Stirling}, {Spencer}, {de la Force},
  {Garrett}, {Fender}, \& {Ogley}}]{stirling2001}
{Stirling}, A.~M., {Spencer}, R.~E., {de la Force}, C.~J., {et~al.} 2001,
  \mnras, 327, 1273

\bibitem[{{Tam} \& {Yang}(2015)}]{TamYang2015}
{Tam}, P.~H.~T. \& {Yang}, Y.-J. 2015, The Astronomer's Telegram, 7327, 1

\bibitem[{{Vila} \& {Romero}(2010)}]{vila2010}
{Vila}, G.~S. \& {Romero}, G.~E. 2010, \mnras, 403, 1457

\bibitem[{{Vila} {et~al.}(2012){Vila}, {Romero}, \& {Casco}}]{vila-etal2012}
{Vila}, G.~S., {Romero}, G.~E., \& {Casco}, N.~A. 2012, \aap, 538, A97

\bibitem[{{Yoon} \& {Heinz}(2015)}]{YoonHeinz2015}
{Yoon}, D. \& {Heinz}, S. 2015, \apj, 801, 55

\bibitem[{{Zdziarski}(2012)}]{zdziarski2012b}
{Zdziarski}, A.~A. 2012, \mnras, 422, 1750

\bibitem[{{Zdziarski} {et~al.}(2012){Zdziarski}, {Lubi{\'n}ski}, \&
  {Sikora}}]{zdziarski2012}
{Zdziarski}, A.~A., {Lubi{\'n}ski}, P., \& {Sikora}, M. 2012, \mnras, 423, 663

\bibitem[{{Zdziarski} {et~al.}(2014){Zdziarski}, {Pjanka}, {Sikora}, \&
  {Stawarz}}]{zdziarski2014}
{Zdziarski}, A.~A., {Pjanka}, P., {Sikora}, M., \& {Stawarz}, {\L}. 2014,
  \mnras, 442, 3243

\bibitem[{{Zhang} {et~al.}(2014){Zhang}, {Xu}, \& {Lu}}]{zhang2014}
{Zhang}, J., {Xu}, B., \& {Lu}, J. 2014, \apj, 788, 143

\end{thebibliography}
\end{document}